\documentclass[10pt]{article}
\usepackage[dvipsnames]{xcolor}
\usepackage{amssymb,amsfonts,amsmath}
\usepackage[]{graphicx}
\usepackage{url}
\usepackage{authblk}
\usepackage[left=2cm,right=3.5cm,top=2cm,bottom=2cm]{geometry}
\usepackage{lineno}

\usepackage[normalem]{ulem}

\definecolor{revcolor}{RGB}{205,0,0}
\newcommand{\revtext}[1]{#1}

\title{\revtext{A bifidelity sparse-grid surrogate model driven by a discrepancy indicator}}
 \author[1]{Matteo Rosellini}
 \author[1]{Filippo Fruzza}
 \author[1]{Alessandro Mariotti}
 \author[1]{Maria Vittoria Salvetti}
 \author[2]{Lorenzo Tamellini}
 \affil[1]{Dipartimento di Ingegneria Civile e Industriale, University of Pisa, Italy}
 \affil[2]{Consiglio Nazionale delle Ricerche - Istituto di Matematica Applicata e Tecnologie Informatiche ``E. Magenes'' (CNR-IMATI), Pavia, Italy}

\begin{document}


\maketitle

\paragraph{Abstract} \revtext{Sparse-grid approximations} based on Lagrange polynomials \revtext{are} one of the staple
methods for approximating functions that are high-dimensional and expensive to evaluate, 
e.g. in the context of PDE-based parametric design exploration. 
\revtext{They are, however, known to have limited \emph{granularity}: refining these sparse-grid approximations
may require multiple additional function evaluations, potentially leading to computational inefficiency. 
To address this issue, we propose a selection criterion for the candidate collocation points introduced by the 
refinement procedure: the function is evaluated only at those points for which a suitable 
(and crucially zero-cost) discrepancy indicator suggests that the additional evaluation would significantly 
change the resulting approximation.}
The remaining collocation points are instead assigned values predicted by the already available \revtext{sparse-grid approximation},
i.e., following a bifidelity approach that reduces costs while preserving accuracy.
The effectiveness of this methodology is demonstrated on several benchmark \revtext{functions} and on an 
engineering application concerning flashback phenomena in hydrogen-fueled perforated burners.

\revtext{sparse-grid approximation};
discrepancy indicators;
complex flow applications;
hydrogen-fueled perforated burners;
flashback phenomena.



\section{Introduction}
The analysis and design of complex engineering systems frequently rely
on high-fidelity computational models - typically, a set of partial differential equations (PDEs) to be solved.
While these models provide accurate predictions, their significant computational cost often makes their use
impractical in scenarios that require \revtext{repeatedly} evaluating the models for different configurations,
such as uncertainty quantification, parametric design exploration, and optimization.
To overcome this limitation, surrogate modeling, i.e., deriving approximations of the original high-fidelity model
that can be evaluated at a significantly reduced computational cost, has emerged as a powerful and widely adopted strategy. 

In this work we focus on parametric design exploration for parametric PDEs,
i.e., our goal is to assess the variability \revtext{of scalar quantities of interest obtained by solving these PDEs as some,
possibly many, of the PDE parameters vary within prescribed design ranges.}
In this context, one typically needs to evaluate the high-fidelity model for several different values of the 
parameters, which is feasible with acceptable computational costs only if good surrogate models  
are available.

\revtext{Among the many surrogate modeling techniques available in the literature 
\cite{gramacy2020surrogates,adcock.brugiapaglia.webster:book,Forrester2008},
we focus here on sparse-grid approximation, which is a well-established technique
\cite{bungartz2004sparse,babuska.nobile.eal:stochastic2,xiu.hesthaven:high,gerstner2003dimension,ma2009adaptive,pflueger:adaptive,
  chkifa:adaptive-taylor,narayan:Leja,rehme.eal:splines,eigel.eal:convergence,feischl2020,piazzola2024algorithm}.
This technique is a collocation approach, i.e., it recasts the problem of building a surrogate model for the quantity of interest
as an interpolation problem over the parametric domain $\Gamma \subset \mathbb{R}^{d}$.  
The PDE solver is seen as an implicit multivariate function $f(\mathbf{x}):\Gamma \rightarrow \mathbb{R}$, 
which gets evaluated for a set of suitably chosen points of the parametric domain, $\mathbf{x}_1,\mathbf{x}_2,\ldots \in \Gamma$. 
The corresponding results $f(\mathbf{x}_1),f(\mathbf{x}_2),\ldots$
are then interpolated by polynomials to obtain the surrogate model; extensions to vector-valued
quantities of interest are straightforward. 
Several kinds of polynomials can be used in this context 
depending on the characteristics of $f$ (such as regularity, 
local features, effective dimensionality): piecewise polynomials \cite{ma2009adaptive,pflueger:adaptive}, splines \cite{rehme.eal:splines}, monomials \cite{chkifa:adaptive-taylor}, Lagrange polynomials \cite{babuska.nobile.eal:stochastic2,xiu.hesthaven:high,eigel.eal:convergence,feischl2020,piazzola2024algorithm}.
In particular, in this work we employ the latter family of polynomials.
This strategy is effective for a large class of applications, thanks to the spectral accuracy of the Lagrange polynomials {for sufficiently smooth functions}. 
Regardless of the choice of polynomial basis, the sparse-grid construction
is known to be relatively robust to the so-called \emph{curse of dimensionality} \cite{bellman1961},
allowing its application to problems involving several tens of parameters, provided that the function to be approximated exhibits appropriate characteristics.}

\revtext{One of the practical limitations of sparse-grid approximation, especially for the Lagrange version considered in this work, is, however,
  its limited \emph{granularity}. Refining a sparse-grid approximation may require multiple additional evaluations of $f$,
  which can be computationally inefficient or even unfeasible in complex, realistic applications where each evaluation entails a high computational cost.
  Certain technical aspects of the construction can be leveraged to reduce the number of additional evaluations of $f$ required at each refinement step,
  such as adding one multi-index at a time to the index set, or using \emph{level-to-knots functions} that increase slowly,
  in combination with suitable families of collocation points, such as Leja points \cite{narayan:Leja}
  (see Section \ref{sec:sg_background} for more details).
  However, these ``tweaks'' are often not informed by the characteristics of $f$ and might yet ultimately lead to a suboptimal allocation of computational effort.
  To tackle this problem, we propose to perform a screening step on the set of collocation
  points at which a function evaluation would be needed to refine a given sparse-grid approximation,
  and to actually evaluate $f$ only at those points $\mathbf{x}^*$ for which a suitable discrepancy indicator 
  suggests that incorporating the function evaluation $f(\mathbf{x}^*)$ would significantly change the resulting approximation.
}

More specifically, we interpret the construction of a \emph{target} \revtext{sparse-grid approximation}
as a refinement step of a \emph{baseline} \revtext{sparse-grid approximation}, which in turn can be seen
as a refinement step of a \emph{coarse} \revtext{sparse-grid approximation}. We further assume nestedness of the three sparse grids.
Upon constructing the baseline \revtext{sparse-grid approximation}, the coarse \revtext{sparse-grid approximation} is also
available, and the difference between the two approximations, relative or absolute, can be used as a zero-cost 
discrepancy indicator, that we evaluate for all the collocation
points at which we still need to evaluate {$f$} to build the target \revtext{sparse-grid approximation}
(\emph{candidate points}). 
The candidates are ranked according to this metric, to select the points for which obtaining the actual function evaluation
is expected to have the greatest impact on the resulting approximation,
subject to a computational budget or a specific {discrepancy} threshold.
The final \revtext{bifidelity surrogate model} is constructed evaluating $f$ at the selected points,
whereas the remaining collocation points are instead assigned values predicted by the already available 
{\emph{baseline}} \revtext{sparse-grid approximation}.
This bifidelity strategy, although admittedly quite simple, 
{can} dramatically reduce the number of expensive function evaluations 
and achieve remarkable results, 
yielding a final surrogate model that closely approximates the accuracy of a fully-resolved, target \revtext{sparse-grid approximation} at a fraction of the computational cost. 

The proposed methodology bears some similarity with other contributions in the literature of surrogate modeling for parametric PDEs.
For instance, \cite{fairbanks:bifid,hampton:bifid} discuss a bifidelity low-rank approach to surrogate modeling construction,
in which the two different fidelities can be either different discretizations of the same PDE, or different
equations for the same physical phenomenon, whereas in our methodology the low-fidelity is the baseline \revtext{sparse-grid approximation}
and the high-fidelity is the full function evaluation;
\cite{Christen:DAMCMC,laloy.etal:pc+mcmc_for_posterior,zacchei.manzoni:MFDA} consider so-called delayed-acceptance algorithms
in which a cheap surrogate model is evaluated prior to deciding whether the expensive function
should be evaluated at the proposal sample of a Markov Chain Monte Carlo sampler;  
\cite{pagani:errorrb,pagani:error2,schiavazzi:error} propose to construct a surrogate model for the error of a reduced basis
(similarly to our zero-cost discrepancy indicator), that can be incorporated in the construction of the reduced basis to improve its accuracy.
Finally, \cite{nobile.eal:adaptive-lognormal} shows how a \revtext{sparse-grid approximation} can be used as a control variate for a Monte Carlo sampling approach.
To the best of our knowledge, this is however the first work in which an approach based on a discrepancy indicator is used purely
in the context of a sparse-grid construction. Moreover, compared to 
\cite{fairbanks:bifid,hampton:bifid,pagani:errorrb,pagani:error2,schiavazzi:error} our discrepancy indicator is truly
zero-cost, and can be implemented with very minimal overhead on pre-existing software.
In particular, {we used a Python code inspired by} the Sparse Grids {Matlab} Kit \cite{piazzola2024algorithm},
that we make publicly available (see Data Availability Statement at the end of this manuscript for details).

\revtext{Our approach also shares some aspects with methodologies for constructing locally adaptive versions of sparse-grid approximation
  and other analogous locally adaptive collocation-based surrogate models.
Such methodologies target problems in which $f$ has localized features, such as sharp gradients, discontinuities, or even pole singularities. 
The literature in this area of research is rich: we mention here  
purely locally adaptive, feature-chasing schemes \cite{ma2009adaptive,pflueger:adaptive,jakeman.archi.xiu:discont,chantrasmi.eal:discontinous}; 
schemes that consider piecewise approximations on Voronoi cells \cite{Rushdi.eal:voronoi}, 
simplices from Delaunay triangulations \cite{witteveen.iaccarino:simplexENO,giovanis.shields:simplexVARcoll,Fuchs.garcke:simplex}, 
hyperrectangular partitions 
\cite{foo.wan.karniadakis:multi-coll,rahman:splinesUQ,galetzka.loukrezis.roemer.eal:hp-adapt-ME-coll},
and finally, schemes that introduce changes of variables and transforms to simplify computations
\cite{colombo.eal:discontinuities,sargsyan.eal:discontinuities,sochala.lemaitre:steep,webster:discontinuities,Gorodetsky.marzouk:discontinuities}.
We emphasize however that the goal of our work is \emph{not} 
to reduce the \emph{error} of a sparse-grid approximation with global polynomials 
when $f$ has local features. Rather, we aim at reducing the \emph{cost} of building such an approximation
whenever $f$ is in principle amenable to it, 
but also very expensive to evaluate as in many practical applications. 
In other words, we are targeting a \emph{different class of problems} from
that tackled by locally adaptive sparse-grid approximations.}

\revtext{From a practical point of view, however, if $f$ exhibits ``moderate'' local features, 
it might nonetheless be more convenient to approximate it 
using our bifidelity strategy rather than a method specifically designed for local adaptivity.
This is because our approximation is of high order and the
discrepancy indicator can be evaluated at zero computational cost, which in turn implies
that the overall cost of our approach is essentially proportional to the number of function evaluations performed.
Conversely, methodologies for local adaptivity are typically of low order
and entail significant algorithmic overhead associated with managing their data structures, which also increases with the number of dimensions.
Moreover, a certain number of function evaluations might be
spent in ``unfruitful'' regions of the parametric domain due to the a posteriori nature of the adaptivity criteria typically employed.
We will discuss this point further when presenting the numerical validation of our approach.
}

The structure of this work is as follows.
In Section \ref{sec:methodology}, we present the theoretical foundations of our {bifidelity surrogate model}, specifically describing
the proposed discrepancy indicator in Section \ref{sec:error-indicator} and designing the final bifidelity surrogate model in Section \ref{sec:surrogate}.
Section \ref{sec:results} is devoted to assessing the performance of the proposed framework.
\revtext{We begin by illustrating the proposed approach for two classic benchmark functions,
  namely a highly oscillatory function (Section \ref{sec:oscillatory}) and the Ishigami function (Section \ref{sec:ishigami}).
  In Section \ref{sec:performance}, extensive validation is carried out using a large set of benchmark functions with different degrees of regularity,
  namely five functions from the Genz collection (oscillatory, product-peak, Gaussian-peak, continuous, and discontinuous functions)
  and the Sobol G-function, with input dimensions ranging from 2 to 11. }
Finally, in Section \ref{sec:experiment}, we demonstrate the applicability of our method to a
real-world scenario involving hydrogen-fueled perforated burners.

\section{Methodology}
\label{sec:methodology}

\subsection{Sparse-grid interpolation: background and notation}
\label{sec:sg_background}

We begin by recalling the essential concepts and notation of sparse-grid interpolation upon which our method is built;
here, we follow closely the setup of \cite{piazzola2024algorithm}, to which we refer readers interested in a more thorough discussion and further bibliographic references.

We consider the problem of approximating a scalar quantity of interest 
$f(\mathbf{x}):\Gamma \subset \mathbb{R}^{d} \rightarrow \mathbb{R}$,
where $\boldsymbol{x} = (x_1,\ldots,x_d)$ denotes a vector of $d$ input parameters
taking values in $\Gamma=\Gamma_1\times\cdots\times\Gamma_d\subset\mathbb{R}^d$, 
and each $\Gamma_n$ is possibly unbounded and characterized by a probability density function (PDF) 
$\rho_n:\Gamma_n\rightarrow\mathbb{R}^+$, $n=1,\ldots,d$.
The specific choice of these PDFs encodes a-priori knowledge/preferences (or uncertainty) on the values of the parameters,
and it could be, e.g., constant, Gaussian or exponential depending on the specific application,
or it may be even available only implicitly as samples obtained from an experiment or a 
Bayesian computation \cite{stuart:acta.bayesian,ghanem:UQbook}.
In this work we assume that $\Gamma = \Gamma_1 \times \cdots \times \Gamma_d$ is a bounded hyper-rectangular domain
and that there are no preferred values for the parameters, i.e., we consider constant PDFs $\rho_n = \frac{1}{|\Gamma_n|}$. 
We further assume that the parameters are independent, such that their joint PDF reads
$\rho(\mathbf{x})=\prod_{n=1}^{d}\rho_n(x_n)$. 

In this setting, \revtext{sparse-grid approximations} provide an efficient collocation-based strategy to construct an interpolant of $f$.
The construction begins with a family of one-dimensional points for each parameter. For
 $n=1,\ldots,d$ and univariate level $i \in \mathbb{N}$, we introduce a set of collocation points
\[
\mathcal{X}^{(n)}_{i} = \left\{ x^{(n)}_{i,j} \right\}_{j=1}^{m(i)},
\]
where $m:\mathbb{N} \rightarrow \mathbb{N} $ is the \revtext{non-decreasing \emph{level-to-knots} function} that prescribes the number of points at each level.
In practice, the points for each parameter $x_n$ are chosen according to $\rho_n$ to improve sampling efficiency and accuracy.
\revtext{In this work, we employ} Clenshaw--Curtis points, 
which are a classical choice whenever the parameters have a constant PDF. 
Clenshaw--Curtis points are defined on a reference interval $[-1,1]$ as
\[
x_{1,1}^{(n)}=0,\qquad
x_{i,j}^{(n)}=
\cos\!\left(\frac{(j-1)\pi}{m(i)-1}\right),
\quad i\geq 2,\quad j=1,\ldots,m(i),
\]
and mapped to $\Gamma_n$ via an affine transformation.
If $m(i)$ is chosen as 
\revtext{
\[
m(i)=
\begin{cases}
     0 & \text{if } i=0\\
     1 & \text{if } i=1\\
     2^{i-1}+1 & \text{if } i\geq 2\\
\end{cases}
\]} 
these points are nested, i.e.\ $\mathcal{X}^{(n)}_{i}\subset\mathcal{X}^{(n)}_{i+1}$, 
allowing collocation points from coarser levels to be systematically reused at finer levels.
This nested structure is a key ingredient of the bifidelity procedure proposed in this work.

Given a multi-index $\boldsymbol{i}=(i_1,\ldots,i_d)\in \mathbb{N}^{d}$, 
a tensor-product grid {is then} defined as the Cartesian product
\[
\mathcal{X}_{\boldsymbol{i}} = \bigotimes_{n=1}^{d} \mathcal{X}^{(n)}_{i_n}.
\]
On this grid, a tensor-product interpolant can be constructed via standard
multivariate Lagrange interpolation. Denoting the associated operator by $\mathcal{U}_{\boldsymbol{i}}$,
the approximation is given by
\[
\mathcal{U}_{\boldsymbol{i}}[f](\boldsymbol{x})
= \sum_{\boldsymbol{x}_{\boldsymbol{i},j}\in \mathcal{X}_{\boldsymbol{i}}}
f(\boldsymbol{x}_{\boldsymbol{i},j})\,\mathcal{L}_{\boldsymbol{i},j}(\boldsymbol{x}),
\]
where $\mathcal{L}_{\boldsymbol{i},j}$ are the standard tensor-product
multivariate Lagrange polynomials, i.e., $\mathcal{L}_{\boldsymbol{i},j}(\boldsymbol{x}_{\boldsymbol{i},k}) = \delta_{j,k}$.
While tensor-product interpolation provides high accuracy, \revtext{the number of points in the underlying grids grows exponentially with the number of dimensions $d$.}
\revtext{Sparse-grid approximations} mitigate this issue by combining a carefully selected set of anisotropic
tensor interpolants through the so-called \emph{combination technique}.  %
In its standard form, given an integer value $w$ (\emph{sparse-grid level})
the \revtext{sparse-grid approximation}  is expressed as a weighted sum
\[
\mathcal{A}_{w}[f](\boldsymbol{x})
= \sum_{\boldsymbol{i}\in \mathcal{I}_{w}} c_{\boldsymbol{i}}\,\mathcal{U}_{\boldsymbol{i}}[f](\boldsymbol{x}),
\]
where the integer coefficients $c_{\boldsymbol{i}}$ are given by the combination technique formula
\[
c_{\boldsymbol{i}} =
\sum_{\substack{\boldsymbol{\delta}\in\{0,1\}^{d} \\ \boldsymbol{i}+\boldsymbol{\delta}\in\mathcal{I}_{w}}}
(-1)^{\lVert \boldsymbol{\delta}\rVert_{1}},
\]
and $\mathcal{I}_{w} = \{\boldsymbol{i} \in \mathbb{N}_{+}^{d} : |\boldsymbol{i}|_1 \leq w + d \}$.
The latter is a classical choice for $\mathcal{I}_{w}$, but other choices are possible, see e.g. \cite{piazzola2024algorithm}.
The associated sparse grid is defined as the union of all tensor grids entering the combination:
\[
\mathcal{X}_{w} = \bigcup_{\boldsymbol{i}\in \mathcal{I}_{w}} \mathcal{X}_{\boldsymbol{i}}.
\]
Note that if nested univariate points are used, $\mathcal{X}_i^{n} \subset \mathcal{X}_{i+1}^{n}$, 
the sparse grids are also nested, $\mathcal{X}_{w-1}\subset\mathcal{X}_{w}$.

\subsection{A zero-cost \revtext{discrepancy} indicator}
\label{sec:error-indicator}

\revtext{As already discussed in the Introduction, in this work we focus on functions $f$ that are potentially very regular (and hence amenable to sparse-grid approximations built using Lagrange polynomials) but also very expensive to evaluate, such that avoiding even a handful of evaluations can result in significant computational savings.}

\revtext{Our approach consists in considering} a simple and relatively large set of candidates \revtext{for function evaluation}, naturally suggested by the nested structure of the sparse-grid construction \revtext{recalled above}, and deciding where $f$ should be evaluated next using a zero-cost discrepancy indicator that relies exclusively on function evaluations that are already available.

More specifically, let us assume that our final goal is to construct a \revtext{sparse-grid approximation} $\mathcal{A}_{w+1}[f]$, which we call the \emph{target} \revtext{sparse-grid approximation}, for some fixed \revtext{level} $w$. Owing to the nestedness of the Clenshaw--Curtis points, this approximation can be computed as a refinement of a \emph{baseline} \revtext{sparse-grid approximation} $\mathcal{A}_{w}[f]$, which we \revtext{assume has already been computed, hence is available}. 
By the same nestedness property, a \emph{coarse} \revtext{sparse-grid approximation} $\mathcal{A}_{w-1}[f]$ \revtext{is also available.
We then introduce the set of candidate points
\begin{equation}
\mathcal{X}_{\mathrm{cand}} = \mathcal{X}_{w+1}\setminus\mathcal{X}_{w},
\label{eq:candidate_set}
\end{equation}
corresponding to the additional evaluations required to construct the target \revtext{sparse-grid approximation}, 
and two pointwise quantities aimed at quantifying the discrepancy between the baseline and coarse \revtext{sparse-grid approximation}s.
The first one is the absolute discrepancy indicator
\begin{equation}
\eta_{\mathrm{abs}}(\mathbf{x}) = \left| \mathcal{A}_{w}[f](\mathbf{x})-\mathcal{A}_{w-1}[f](\mathbf{x})\right|.
\label{eq:eta_abs}
\end{equation}
The second is a symmetric, regularized relative discrepancy indicator,
\begin{equation}
\eta_{\mathrm{rel}}(\mathbf{x})
=\frac{2\left|\mathcal{A}_{w}[f](\mathbf{x})-\mathcal{A}_{w-1}[f](\mathbf{x})\right|}{
\left|\mathcal{A}_{w}[f](\mathbf{x})\right|+\left|\mathcal{A}_{w-1}[f](\mathbf{x})\right|+\varepsilon S},
\label{eq:eta_indicator}
\end{equation}
where $\varepsilon>0$ is a dimensionless regularization parameter
(e.g., $\varepsilon=10^{-6}$) and $S$ is a reference scale, i.e. 
\begin{equation}
S=\max_{\mathbf{x}\in\mathcal{X}_{\mathrm{cand}}}
\max\left\{
\left|\mathcal{A}_{w}[f](\mathbf{x})\right|,
\left|\mathcal{A}_{w-1}[f](\mathbf{x})\right|
\right\}.
\label{eq:indicator_scale}
\end{equation}}

\revtext{Crucially, both indicators can be evaluated without additional evaluations of $f$.
  The absolute indicator measures changes in the physical units of the output and is suitable when absolute discrepancies are of primary interest.
  However, it may prioritize regions with large response magnitudes, even when their relative changes are small.
  Conversely, the relative indicator is dimensionless and invariant under multiplicative rescaling of the output,
  allowing relative changes to be compared across regions with different response magnitudes.
  Its main limitation is that small absolute discrepancies may receive high priority where the response approaches zero.
  The regularization parameter $\varepsilon$ mitigates this behavior, although its choice influences the indicator near zero.}

\revtext{After choosing either indicator, denoted by $\eta$, 
  candidate points are ranked in descending order according to their indicator values,
  and all candidates such that $\eta(\mathbf{x}) \geq \eta_{\mathrm{cut}}$ for a given 
  $\eta_{\mathrm{cut}}$ are selected for evaluation. We consider two approaches to define the cutoff value $\eta_{\mathrm{cut}}$:
\begin{enumerate}
\item \emph{Relative threshold.} It is a \emph{direct} definition strategy:
  for a prescribed threshold $\tau\in(0,1]$, the cutoff value is defined as
  \begin{equation}
    \eta_{\mathrm{cut}} = \tau\eta_{\max},
    \label{eq:threshold}
  \end{equation}
  where $\eta_{\max}$ is the maximum indicator value over the candidate set. 
  This strategy adapts the number of evaluations to the indicator distribution, avoiding candidates with indicator values that are small relative to the maximum value.
  However, the computational cost cannot be prescribed in advance, and the selection may be sensitive to the threshold and to anomalously large maximum values.
\item \emph{Fractional budget.} It is an \emph{indirect} definition strategy: we prescribe the fraction of candidate points to be 
  selected, i.e., we choose $q\in[0,1]$ and allow
  \begin{equation}\label{eq:Nq}
    N_{q} = \left\lfloor q\left|\mathcal{X}_{\mathrm{cand}}\right| \right\rfloor
  \end{equation} 
  function evaluations. The cutoff value $\eta_{\mathrm{cut}}$ is then defined as 
  the indicator value of the $N_{q}$-th highest-ranked candidate.
  This strategy provides direct control over the nominal computational cost.
  However, it does not adapt to the indicator distribution: a small budget may exclude relevant candidates,
  whereas a large budget may include points with negligible indicator values.
\end{enumerate}
}
\revtext{Regardless of the chosen approach to define $\eta_{\mathrm{cut}}$, a strict cutoff strategy could end up excluding from 
  the selection candidates with equal or nearly equal indicator values, which might lead to a degradation of the performance of the proposed methodology.
  To prevent this, we introduce a cutoff-dependent tolerance
\begin{equation}
\delta_{\mathrm{eq}}
=
10^{-3} \eta_{\mathrm{cut}}
\label{eq:selection_tolerance}
\end{equation}
and finally define the selected set as
\begin{equation}
\mathcal{X}_{\mathrm{sel}}
=
\left\{
\mathbf{x}\in\mathcal{X}_{\mathrm{cand}}:
\eta(\mathbf{x})
\geq
\eta_{\mathrm{cut}}-\delta_{\mathrm{eq}}
\right\}.
\label{eq:selected_set}
\end{equation}
Note that this means that if the \emph{fractional budget} strategy was employed, the final number
of function evaluations might exceed the nominally prescribed number $N_q$. }

\subsection{Design of the final surrogate model}
\label{sec:surrogate}

Once the set of points marked for evaluation $\mathcal{X}_{\mathrm{sel}}$ is available, we can construct the final surrogate model. Crucially,
this is \emph{not} the target \revtext{sparse-grid approximation} $\mathcal{A}_{w+1}[f]$, but rather a bifidelity
surrogate model $\mathcal{B}[f] \approx \mathcal{A}_{w+1}$, in which the points in $\mathcal{X}_{\mathrm{cand}} \setminus \mathcal{X}_{\mathrm{sel}}$
will be assigned the value of $\mathcal{A}_{w}[f]$ rather than $f$.
Equivalently, $\mathcal{B}[f]$ can be seen as the \revtext{sparse-grid approximation} of a generic continuous function $g$ whose value
in the collocation points $\mathcal{X}_{w+1}$  is defined as follows:
\begin{equation}
  \mathcal{B}[f] =
  \mathcal{A}_{w+1}[g], \quad
 g(\mathbf{x}_k) = 
  \begin{cases} 
    f(\mathbf{x}_k) & \text{if } \mathbf{x}_k \in \mathcal{X}_{\mathrm{sel}} \cup  \mathcal{X}_{w} \\
    \mathcal{A}_{w}[f](\mathbf{x}_k) & \text{if } \mathbf{x}_k \in \mathcal{X}_{\mathrm{cand}} \setminus \mathcal{X}_{\mathrm{sel}}
  \end{cases}.
  \label{eq:hybrid_g}
\end{equation}
We conclude this section by remarking that it is possible to rewrite $\mathcal{B}[f]$  to highlight its \emph{incremental} nature by rewriting 
$g = \mathcal{A}_{w}[f]+ \zeta$,  with 
\[
\zeta(\mathbf{x}_k) = 
  \begin{cases} 
    f(\mathbf{x}_k) - \mathcal{A}_{w}[f](\mathbf{x}_k) & \text{if } \mathbf{x}_k \in \mathcal{X}_{\mathrm{sel}} \cup  \mathcal{X}_{w} \\
    0 & \text{if } \mathbf{x}_k \in \mathcal{X}_{\mathrm{cand}} \setminus \mathcal{X}_{\mathrm{sel}}
  \end{cases}
\]
and then noting that 
\[
  \mathcal{B}[f] =
  \mathcal{A}_{w+1}[g] =
  \mathcal{A}_{w+1}\Big[\mathcal{A}_{w}[f] + \zeta\Big] =
  \mathcal{A}_{w}[f] + \mathcal{A}_{w+1}[\zeta],
\]
where in the last equality we have used the linearity of the interpolation operators and the fact that $\mathcal{A}_{w}[f]$ is a polynomial that can be interpolated exactly by $\mathcal{A}_{w+1}$ due to the nestedness of the corresponding grids. The rightmost expression shows how $\mathcal{B}[f]$ can be obtained by adding the interpolant of a suitable correction/increment term to the baseline \revtext{sparse-grid approximation} $\mathcal{A}_{w} [f]$, in the spirit of \cite{pagani:errorrb,pagani:error2}.

\section{Results}
\label{sec:results}
To validate the performance of our approach, we first \revtext{discuss in depth its application to a couple of illustrative examples} (Section \ref{sec:analytical}).
We then move \revtext{to an extensive numerical campaign over a set of benchmark functions with different regularity (Section \ref{sec:performance}), and finally}
to a practical engineering case study (Section \ref{sec:experiment}). For the illustrative examples and the burner case study, we consider baseline and target sparse-grid approximations at levels $w=2$ and $w=3$, respectively. This setting reflects realistic engineering constraints, where the high computational cost of simulations allows only for very few full model evaluations; this is indeed exemplified in Section \ref{sec:experiment}. Note that in this sense the dimension-adaptive \revtext{sparse-grid approximations}  \cite{gerstner2003dimension,piazzola2024algorithm} are not a viable option to derive a baseline \revtext{sparse-grid approximation}, since they are also a posteriori algorithms that invest a non-negligible cost at the beginning of the computation to assess what parameters are most influential and hence worth refining. 

\subsection{\revtext{Illustrative examples}}
\label{sec:analytical}
\revtext{The chosen benchmark functions are a continuously differentiable oscillatory function introducing directional gradients of different intensity, and the Ishigami function \cite{ishigami1990importance}.}


\subsubsection{Oscillatory function}
\label{sec:oscillatory}

\revtext{As a first example,} we consider the smooth oscillatory function 
\[ 
f(\mathbf{x}) = \sum_{i=1}^{d} \cos(\theta_i x_i), \qquad \mathbf{x}\in[0,1]^d,\quad d=4, 
\] with $\boldsymbol{\theta} = \left[1.5\pi,\;3.0\pi,\;0.5\pi,\;4.5\pi\right]$.
The selected frequencies generate directional variations of different scales, ranging from low-frequency behavior along $x_3$ to higher-frequency oscillations along $x_2$ and $x_4$. 

\begin{figure}[ht]
    \centering
    \includegraphics[width=0.48\textwidth]{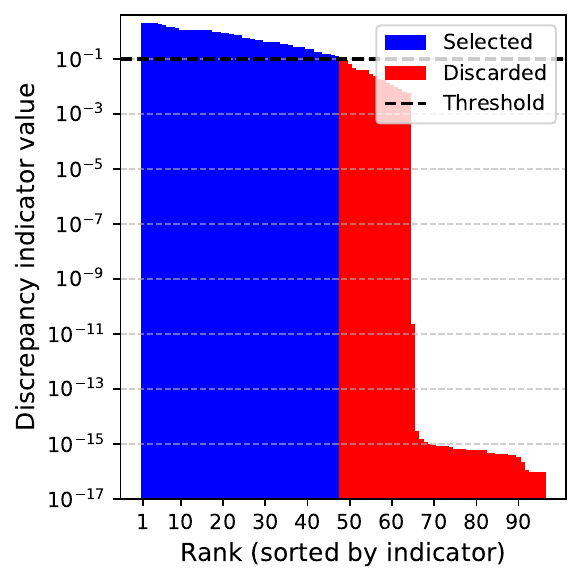}
    \includegraphics[width=0.48\textwidth]{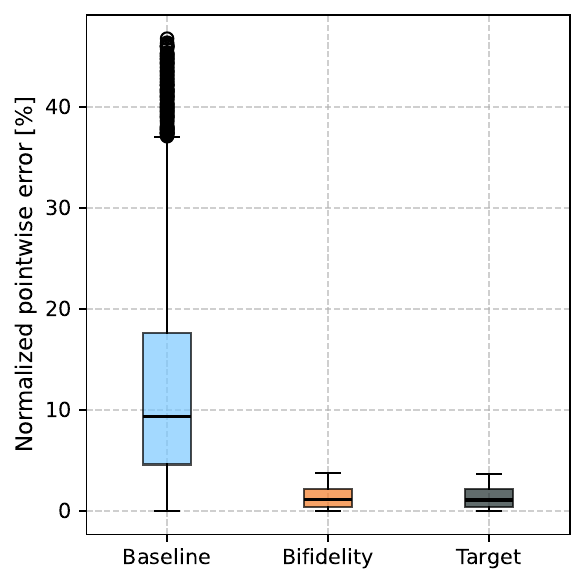}
    \caption{Left: discrepancy indicator for the oscillatory function. Right: boxplots of the percentage normalized error
      of the three approximations (baseline \revtext{sparse-grid approximation}, bifidelity surrogate model, target \revtext{sparse-grid approximation})
      of the oscillatory function.}
    \label{fig:oscillatory-error-combined}
\end{figure}

Figure \ref{fig:oscillatory-error-combined} (left) shows the values of the relative discrepancy indicator $\eta$ 
for the collocation points in $\mathcal{X}_{\mathrm{cand}}$, sorted decreasingly.
Applying the threshold-based \revtext{definition of $\eta_{\mathrm{cut}}$ with $\tau=5\%$, see Equation \eqref{eq:threshold},}
results in the selection of $\lvert\mathcal{X}_{\mathrm{sel}}\rvert = 47$ points out of 96, that are then used to construct the bifidelity surrogate model.
Figure \ref{fig:oscillatory-error-combined} (right) presents the boxplot of the 
percentage normalized pointwise error for the baseline \revtext{sparse-grid approximation} 
$\mathcal{A}_{2}[f]$, the bifidelity surrogate model $\mathcal{B}[f]$, and the target \revtext{sparse-grid approximation} $\mathcal{A}_{3}[f]$, 
evaluated over \revtext{a set of 200 randomly selected test points. More precisely, 
upon denoting the set of test points by $\mathcal{E} \subset \Gamma$, 
the error for each $\mathbf{x} \in \mathcal{E}$ is computed as
\begin{equation}
\label{eq:errore_ranking}
\varepsilon(\mathbf{x}) =
100 \times 
\frac{\lvert \mathcal{C}[f](\mathbf{x}) - f(\mathbf{x}) \rvert}{\max_{\mathbf{z} \in \mathcal{E}} f(\mathbf{z}) - \min_{\mathbf{z} \in \mathcal{E}} f(\mathbf{z})},
\end{equation}
where $\mathcal{C}[f]$ is a placeholder for any of $\mathcal{A}_{2}[f]$, $\mathcal{B}[f]$, 
and $\mathcal{A}_{3}[f]$.}
The baseline sparse-grid approximation exhibits a broader error distribution, with several points reaching relatively large errors. In contrast, the bifidelity surrogate model shows a much more concentrated distribution, maintaining error levels below 10\% at all evaluated points, and the target sparse-grid approximation performs only marginally better.

\begin{figure}[t]
    \centering
    \includegraphics[width=\linewidth]{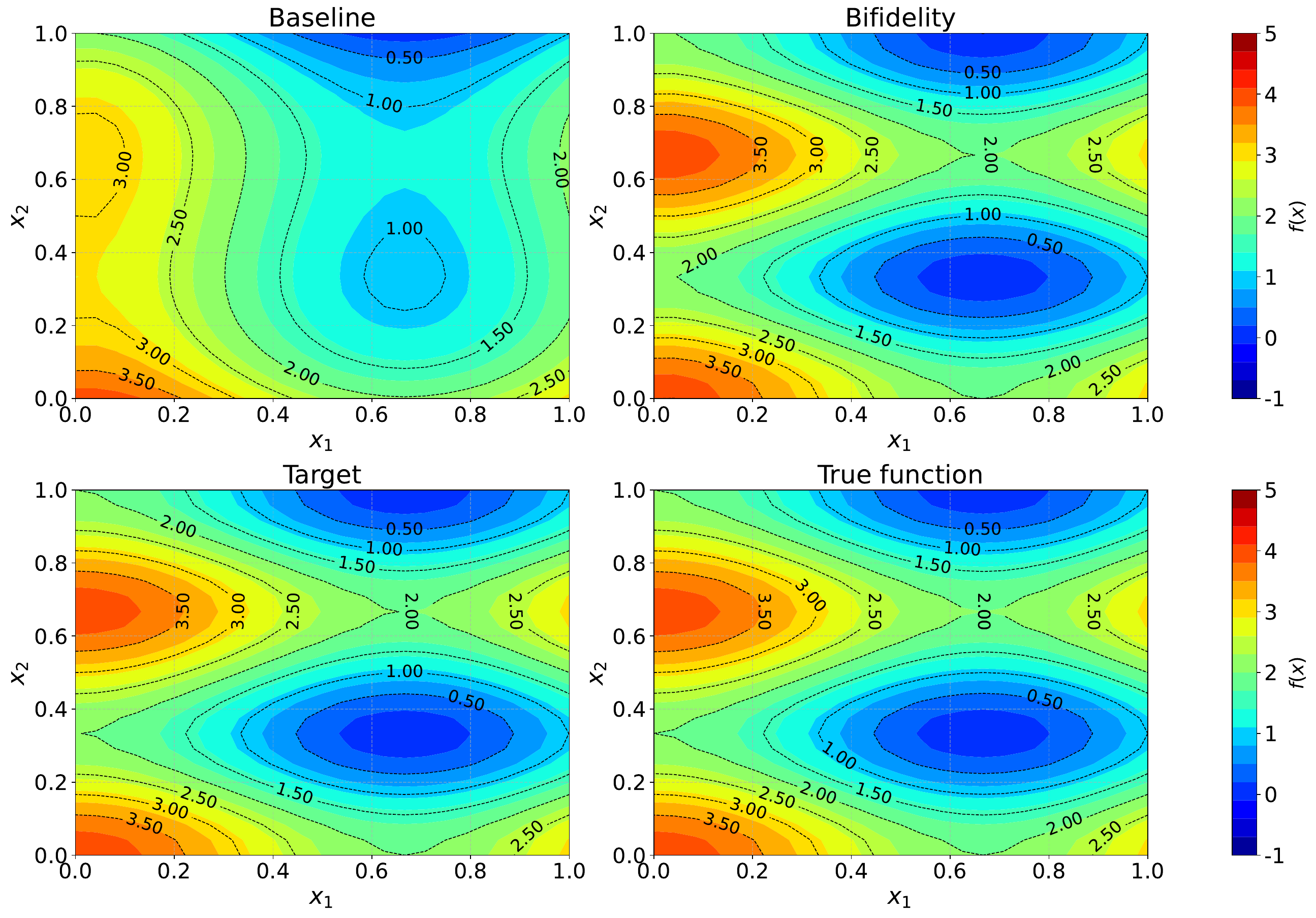}
    \caption{Baseline \revtext{sparse-grid approximation}, bifidelity surrogate model, and target \revtext{sparse-grid approximation}
      compared against the true oscillatory function at $x_3=x_4=0$}
    \label{fig:oscillatory-response1}
\end{figure}
\begin{figure}[t]
    \centering
    \includegraphics[width=0.8\linewidth]{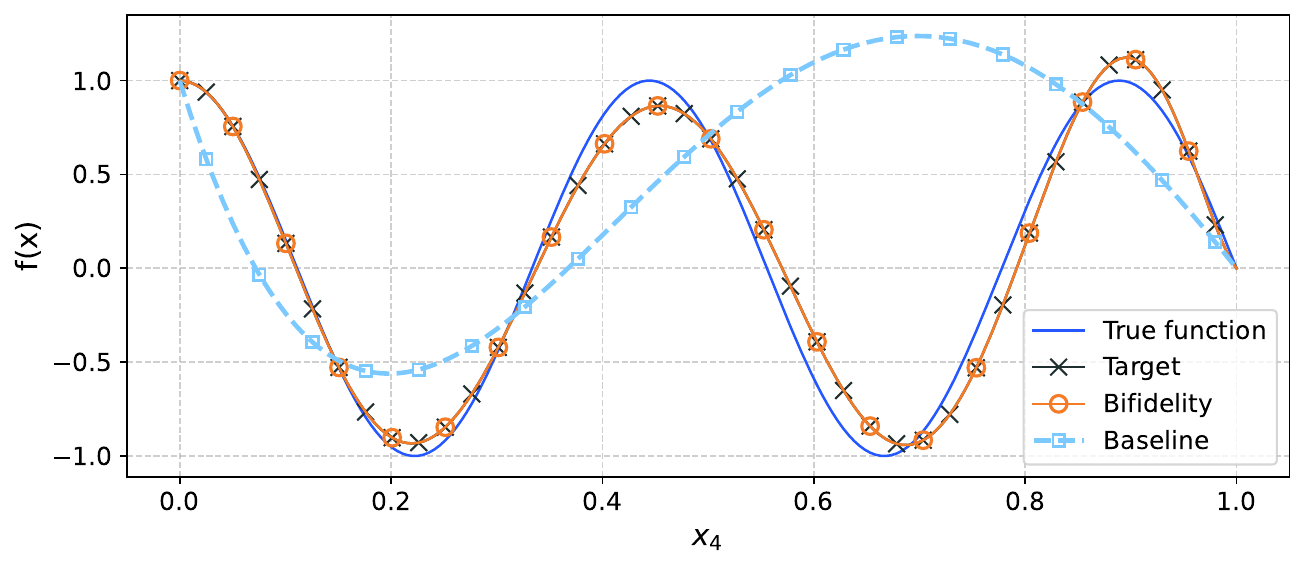}
    \caption{Baseline \revtext{sparse-grid approximation}, bifidelity surrogate model, and target \revtext{sparse-grid approximation}
      compared against the true oscillatory function at $x_1=x_2=x_3=0.5$.}
    \label{fig:oscillatory-errorline1}
\end{figure}

\revtext{We also provide a visual comparison of }
the three approximations (baseline \revtext{sparse-grid approximation}, bifidelity surrogate model, and target \revtext{sparse-grid approximation}) against the true function both in a two-dimensional cross-section (Figure \ref{fig:oscillatory-response1}, fixing $x_3 = x_4 = 0$), as well as in a one-dimensional cross-section (Figure \ref{fig:oscillatory-errorline1}, fixing $x_1 = x_2 = x_3 = 0.5$). It is visually confirmed that the baseline sparse-grid approximation provides nothing more than a coarse approximation of \revtext{the true} function. The \revtext{one-dimensional cross-section} in Figure \ref{fig:oscillatory-errorline1} shows in particular that even the number of peaks is not properly captured. Conversely, the target sparse-grid approximation and bifidelity surrogate model provide a very reasonable approximation of the true function and are again nearly identical, which is particularly remarkable since, in this test, the bifidelity surrogate model uses only about a half of the function evaluations of the target \revtext{sparse-grid approximation} (47 out of 96).


\begin{figure}[t]
    \centering
    \includegraphics[width=0.48\textwidth]{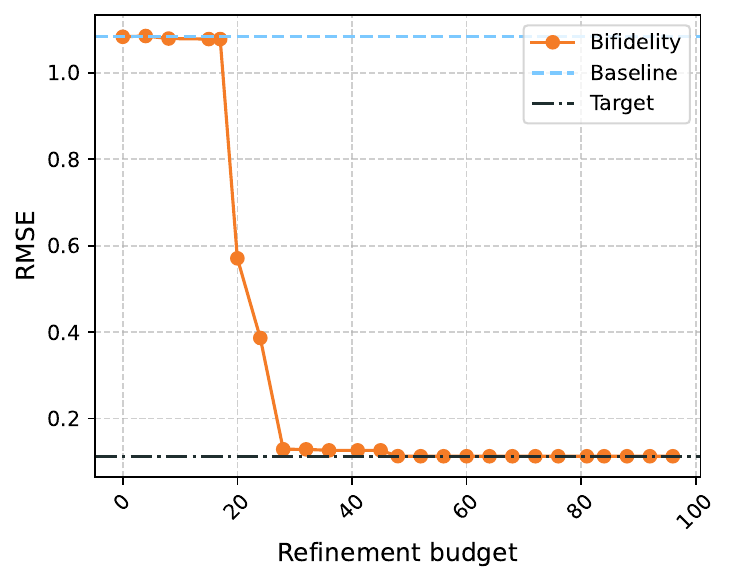}
    \includegraphics[width=0.48\textwidth]{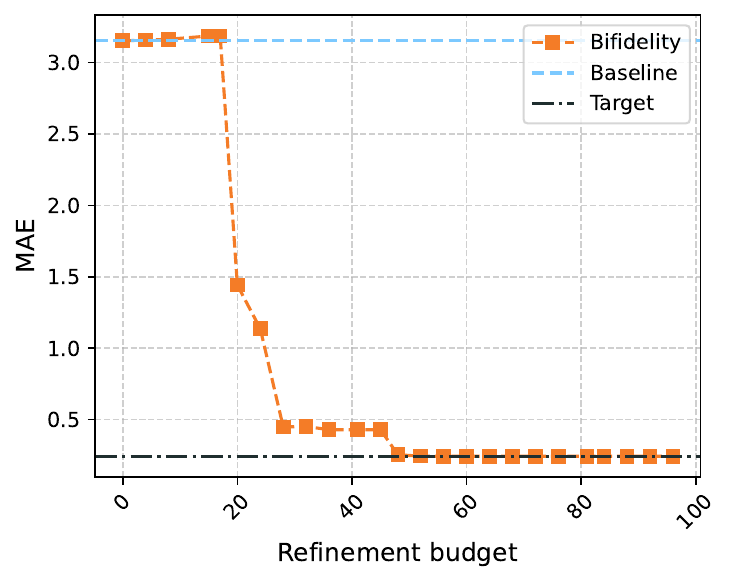}
    \caption{Convergence of RMSE (left) and MAE (right) for the oscillatory function.}
    \label{fig:oscillatory-convergence}
\end{figure}

\revtext{Finally, in Figure \ref{fig:oscillatory-convergence}, we investigate the sensitivity to the number of points included in $\mathcal{X}_{sel}$, i.e., 
we monitor how the approximation quality of $\mathcal{B}[f]$ evolves as the computational budget increases.
This test is crucial since it directly checks the ability of our bifidelity strategy to actually
recover the accuracy of the target \revtext{sparse-grid approximation} at a limited computational cost,
which is the whole point of our approach. 
In detail, we construct a sequence of bifidelity surrogate models, 
each incorporating an increasing number of refinement points starting from
$\mathcal{X}_{sel} = \emptyset$, i.e., the bifidelity surrogate model is identical 
to the baseline \revtext{sparse-grid approximation} ($\mathcal{B}[f] = \mathcal{A}_2[f]$),
until $\mathcal{X}_{sel} = \mathcal{X}_{cand}$, i.e. the bifidelity surrogate model is identical to the target sparse-grid approximation ($\mathcal{B}[f] = \mathcal{A}_3[f]$).
We introduce again a test set of $M=200$ randomly distributed points $\mathcal{E} \subset \Gamma$,
and for each bifidelity surrogate model we evaluate its root mean square error (RMSE): 
\begin{equation}
\label{eq:rmse}
\text{RMSE} = \sqrt{\frac{1}{M}\sum_{\mathbf{x} \in \mathcal{E}} \left( f(\mathbf{x}) - \mathcal{B}[f](\mathbf{x}) \right)^2},
\end{equation}
and its maximum absolute error (MAE): 
\begin{equation}
\label{eq:mae}
\text{MAE} = \max_{\mathbf{x} \in \mathcal{E}} \left| f(\mathbf{x}) - \mathcal{B}[f](\mathbf{x}) \right|.
\end{equation}
Figure \ref{fig:oscillatory-convergence} shows that both errors decay ``almost monotonically'' 
as the number of evaluations increases, and in particular we observe a convergence plateau beyond which further point additions result in negligible error reduction.
This confirms that in general a bifidelity approach is enough to recover full accuracy of the target \revtext{sparse-grid approximation}, i.e., the target \revtext{sparse-grid approximation} has in a way some 
``redundancy'' property. 
In practice, the amount of computational savings depends of course on the specific function
as well as on the ability of the discrepancy indicator $\eta$ to mimic the decay of the actual error.}



\subsubsection{The Ishigami Function}
\label{sec:ishigami}

\begin{figure}[t]
    \centering
    \includegraphics[width=0.48\textwidth]{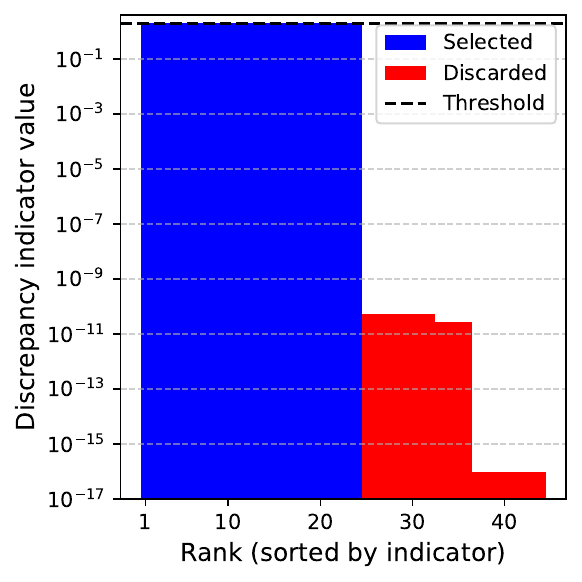}
    \includegraphics[width=0.48\textwidth]{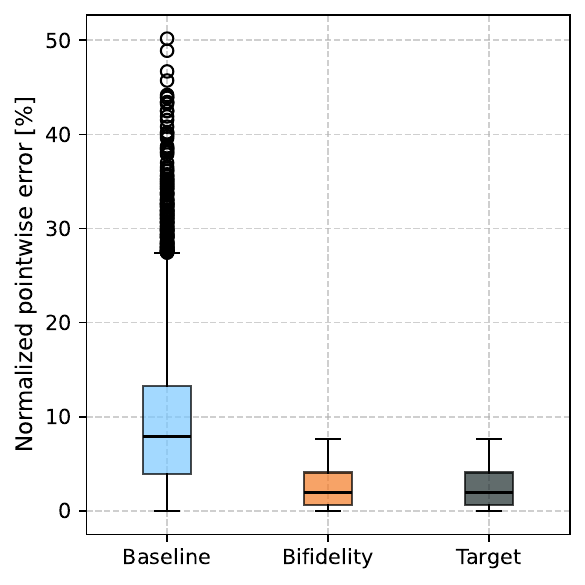}
    \caption{Left: discrepancy indicator for the Ishigami function. Right: boxplots of the percentage normalized error of the three approximations (baseline \revtext{sparse-grid approximation}, bifidelity surrogate model, target \revtext{sparse-grid approximation}) of the Ishigami function.}
    \label{fig:ishigami-error-combined}
\end{figure}

The Ishigami function  \cite{ishigami1990importance} is a non-linear, non-monotonic benchmark ($d=3$),
characterized by a strong interaction between its first and third variables. It is defined as:
\[
    f(\mathbf{x}) = \sin(x_1) + a \sin^2(x_2) + b x_3^4 \sin(x_1),
\]
whose domain is $\mathbf{x} \in [-\pi, \pi]^3$, and the coefficients are $a=7$ and $b=0.1$.
Figure \ref{fig:ishigami-error-combined} (left) shows the values of the relative discrepancy indicator $\eta$ 
for the collocation points in $\mathcal{X}_{\mathrm{cand}}$, sorted decreasingly.
In this test the {discrepancy} indicator 
is \revtext{constant} for most of the points, 
meaning that a ranking of points is not really possible. 
This is due to the fact that the value of the Ishigami function 
at the 7 collocation points used to build the  coarse sparse-grid approximation $\mathcal{A}_{1}[f]$ is always zero:
therefore, $\mathcal{A}_{1}[f]=0$ on the entire domain,
and the {discrepancy indicator has value} $\eta(\mathbf{x})=2$ 
for most points in $\mathcal{X}_{\mathrm{cand}}$ (see Equation \eqref{eq:eta_indicator}). 
The only exceptions are those points at which
also $\mathcal{A}_{2}[f]=0$ and hence \revtext{$\eta$ is numerically zero.} 
Including in $\mathcal{X}_{\mathrm{sel}}$ all the points with $\eta(\mathbf{x})=2$ 
results in a set $\mathcal{X}_{\mathrm{sel}}$ with \revtext{24} points out of 44, 
that are then used to construct the bifidelity surrogate model.

\revtext{As in the previous analysis,} Figure \ref{fig:ishigami-error-combined} (right) 
presents the boxplot of the 
\revtext{percentage normalized pointwise} error for the baseline \revtext{sparse-grid approximation} 
$\mathcal{A}_{2}[f]$, the bifidelity surrogate model $\mathcal{B}[f]$, and the target \revtext{sparse-grid approximation} 
$\mathcal{A}_{3}[f]$, evaluated over \revtext{a set of 200 randomly selected test points} 
\revtext{and defined as in Equation \eqref{eq:errore_ranking}}.
The baseline sparse-grid approximation exhibits \revtext{}{also in this case} a broad error distribution, 
with large \revtext{interquartile} range, heavy tails and large errors reaching up to 50\%. 
In contrast, the bifidelity surrogate model shows a sharply concentrated distribution around lower error values, 
with no instances exceeding 10\% and overall performance identical to that of the target sparse-grid approximation, 
confirming that the bifidelity surrogate model can effectively reproduce the accuracy of a 
more refined \revtext{sparse-grid approximation} at a fraction of the cost.

\begin{figure}[t]
    \centering
    \includegraphics[width=\linewidth]{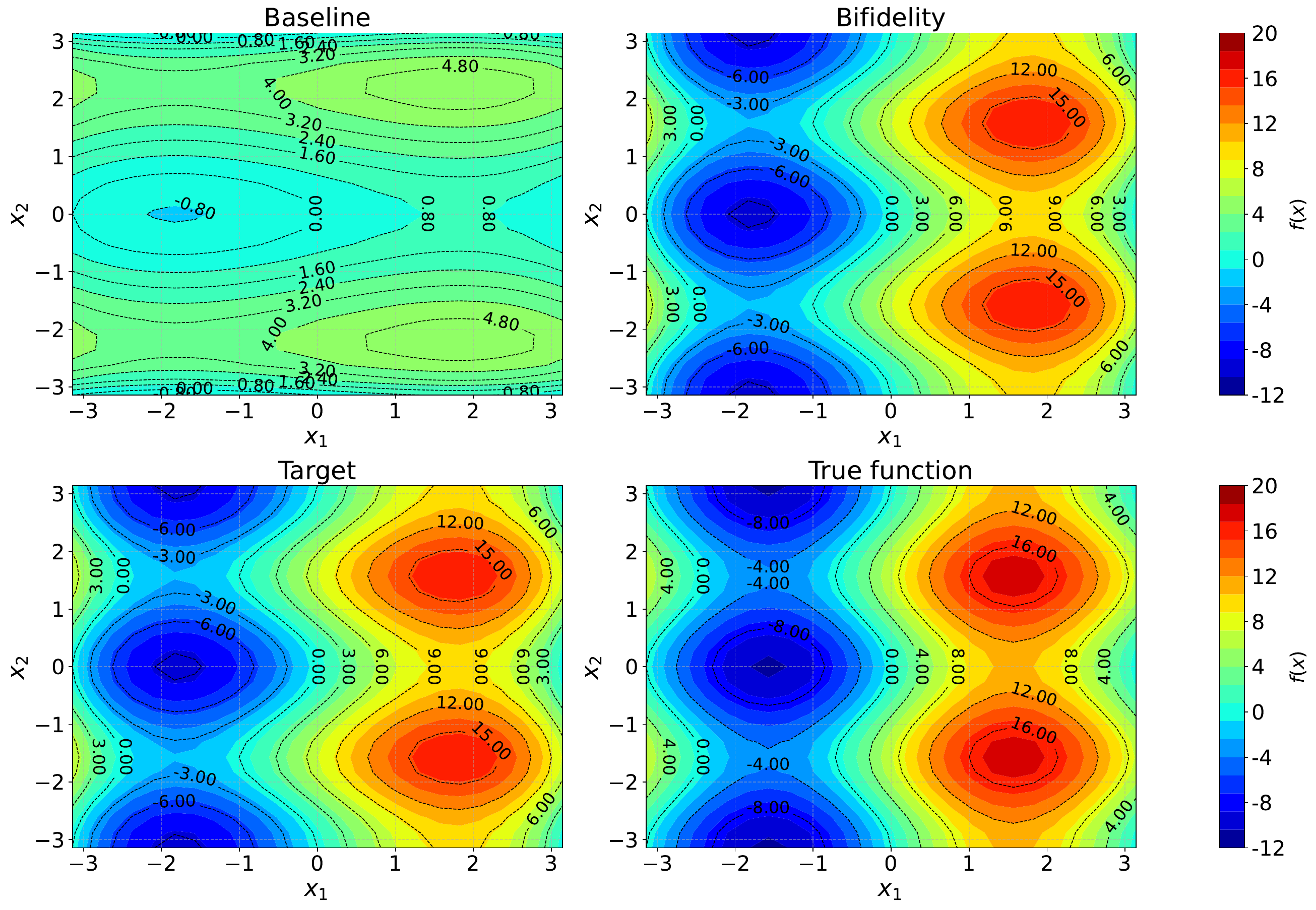}
    \caption{Baseline \revtext{sparse-grid approximation}, bifidelity surrogate model, and target \revtext{sparse-grid approximation}
      compared against the true Ishigami function at $x_3=\pi$.}
    \label{fig:ishigami-response2}
\end{figure}

\begin{figure}[t]
    \centering
    \includegraphics[width=0.8\linewidth]{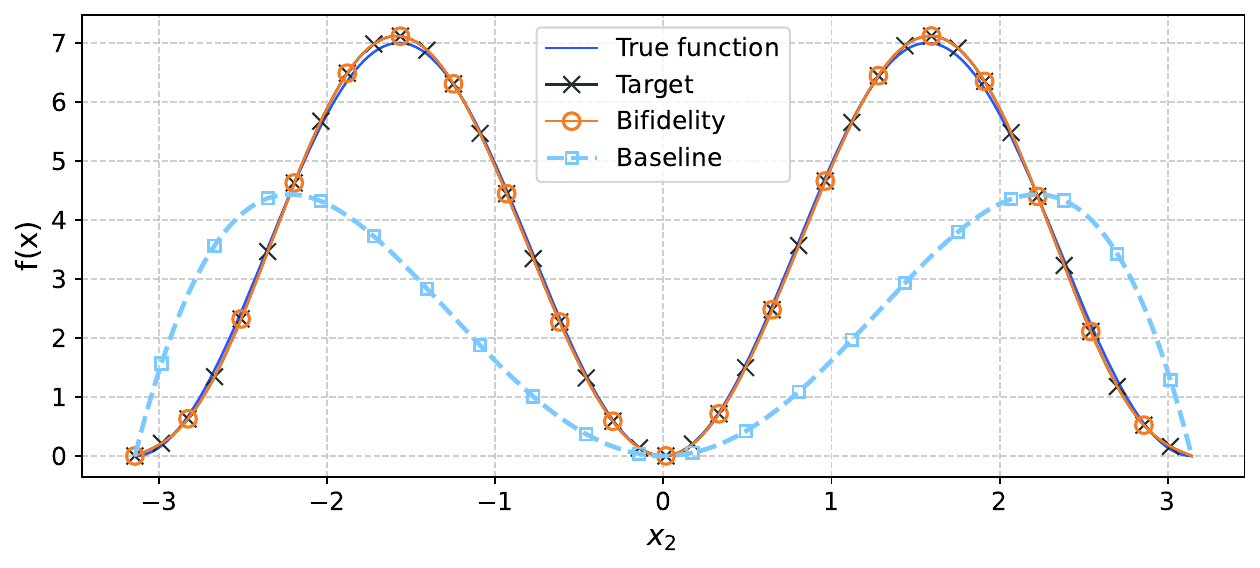}
    \caption{Baseline \revtext{sparse-grid approximation}, bifidelity surrogate model, and target \revtext{sparse-grid approximation}
      compared against the true Ishigami function  at $x_1=x_3=\pi$}
    \label{fig:ishigami-errorline1}
\end{figure}

Figures \ref{fig:ishigami-response2} and \ref{fig:ishigami-errorline1} \revtext{then compare visually}
the three approximations (baseline \revtext{sparse-grid approximation}, bifidelity surrogate model, and target \revtext{sparse-grid approximation})
against the true Ishigami function.  
The former \revtext{figure} plots the cross-section \revtext{in the} $(x_1, x_2)$ plane, fixing $x_3 = \pi$, 
while the latter one plots the cross-section on the line defined by $x_1=x_3 = \pi$.
{These} figures confirm that the baseline sparse-grid approximation merely provides a coarse approximation of the overall trend of the Ishigami function. It fails to capture both the location and the amplitude of the sine-induced peaks, resulting in a visibly smeared approximation.
Conversely, the target sparse-grid approximation and bifidelity surrogate model are nearly identical; moreover, in this case both essentially capture the entire landscape of the Ishigami function.
Once again, the bifidelity surrogate model reaches the accuracy of the target \revtext{sparse-grid approximation} with a smaller computational cost (\revtext{24} function evaluations instead of 44, corresponding to a sizable 
\revtext{45\%} computational saving).

\begin{figure}[t]
    \centering
    \includegraphics[width=0.48\textwidth]{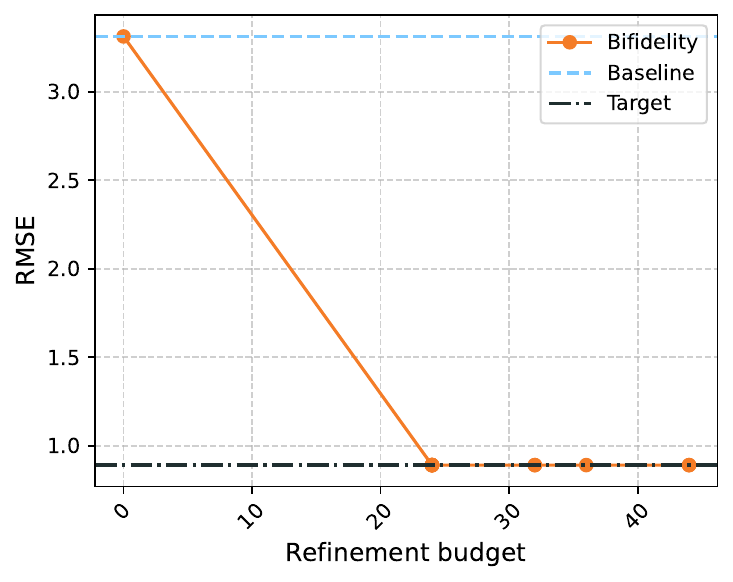}
    \includegraphics[width=0.48\textwidth]{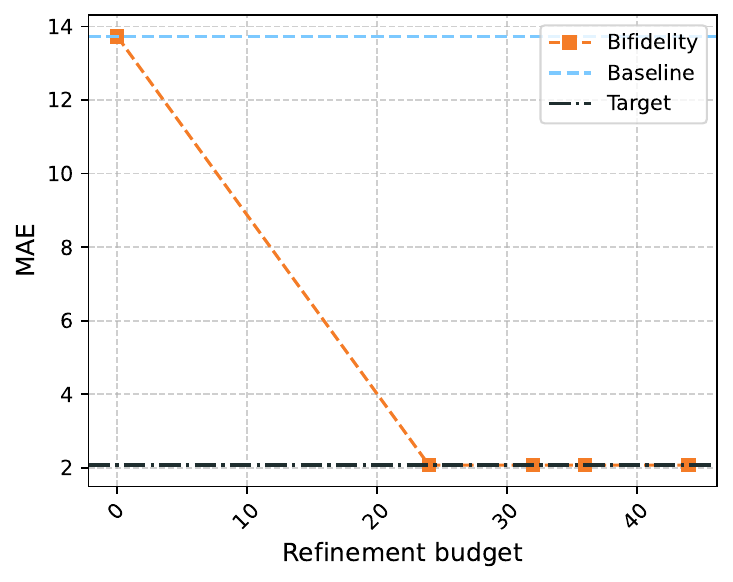}
    \caption{Convergence of RMSE (left) and MAE (right) for the \revtext{Ishigami} function.}
    \label{fig:ishigami-convergence}
\end{figure}

\revtext{Finally, we monitor how the approximation quality of the bifidelity approach evolves as the computational budget increases.}
\revtext{The results, reported in Figure~\ref{fig:ishigami-convergence}, show also for this test an almost monotonic error decay as the number of high-fidelity evaluations increases and a clear convergence plateau, beyond which additional points provide negligible improvements. In this pathological example, $\eta$ is actually a very poor indicator of the error decay, as it takes only two values (the maximum value 2 and machine-zero) and remains constant for 24 out of the 44 candidate points. Nevertheless, owing to the tolerance-based selection strategy introduced in Equation~\eqref{eq:selection_tolerance}, the method is still able to identify the relevant candidate set. In fact, all points associated with the maximum indicator value are selected simultaneously, so that the first enrichment step already includes the 24 candidates responsible for most of the approximation improvement. As a result, while $\eta$ provides little information for ranking points within this group, it still allows the method to reach the convergence plateau, avoiding a fraction of the high-fidelity computations while yielding an approximation that remains competitive with the target sparse-grid approximation.}



\subsection{\revtext{Efficiency of the bifidelity construction}}
\label{sec:performance}

\revtext{The aim of this section is to extensively test the efficiency of the bifidelity construction
  in recovering the accuracy of the target sparse-grid approximation at a fraction of the computational cost.
  To this end, we consider a set of six test functions of varying smoothness (hence extending even outside
  the intended range of applicability of our methodology) for different values of $d$, 
  and for each case monitor how the approximation quality of the bifidelity approach evolves as the computational
  budget increases; in other words, we show for each case a plot analogous to those in Figures \ref{fig:oscillatory-convergence} and \ref{fig:ishigami-convergence}.
  In doing so, we also compare our method with a standard locally adaptive 
  piecewise-linear sparse-grid approximation (in the implementation provided by PyApprox~\cite{jakeman2023pyapprox}),
  and draw some conclusions in this sense.
  The results that we show were obtained employing Clenshaw--Curtis points in the sparse-grid construction;
  we also repeated the same experiments with Leja points \cite{narayan:Leja} and find qualitatively identical results. 
}

\subsubsection{\revtext{Benchmark setup and test functions}}\label{sec:performance-setup}
\revtext{For this benchmark campaign, the baseline and target sparse-grid
approximations are constructed at levels $w=3$ and $w=4$, respectively.}
\revtext{We consider six benchmark functions defined on $[0,1]^d$ with $d=2,5,8,11$: 
five functions from the Genz library \cite{barthelmann.novak.ritter:high}
(oscillatory, product-peak, Gaussian, continuous, discontinuous), 
and the Sobol G-function \cite{sobol1990sensitivity}. They are defined as
\[
\begin{aligned}
& f_{\mathrm{osc}}(\boldsymbol{x})  = \cos \left(\pi+\sum_{i=1}^{d} c_i x_i\right), &
& f_{\mathrm{prod}}(\boldsymbol{x})  = \prod_{i=1}^{d} \left( c_i^{-2}+\left(x_i-\frac{1}{2}\right)^2\right)^{-1},\\
& f_{\mathrm{Gauss}}(\boldsymbol{x}) =\exp \left(-\sum_{i=1}^{d}c_i^2\left(x_i-\frac{1}{2}\right)^2\right), &
& f_{\mathrm{Sobol}}(\boldsymbol{x}) =\prod_{i=1}^{d}\frac{|4x_i-2|+c_i}{1+c_i},\\
& f_{C^0}(\boldsymbol{x})         =\exp \left(-\sum_{i=1}^{d}c_i\left|x_i-\frac{1}{2}\right|\right), &
& f_{\mathrm{disc}}(\boldsymbol{x})= 
\begin{cases}
\displaystyle \exp \left(\sum_{i=1}^{d} c_i x_i\right), & \displaystyle x_1\leq \frac{1}{2}\ \text{and}\ x_2\leq \frac{1}{2},\\[1mm]
0, & \text{otherwise}.
\end{cases}
\end{aligned}
\]
For the Genz functions, the coefficients are prescribed using the anisotropic profile $c_i=i^{-2}/\sum_{j=1}^{d}j^{-2}$, $i=1,\ldots,d$,
such that $\sum_{i=1}^{d}c_i=1$. For the Sobol G-function, instead, $c_i=(i-1)/2$, $i=1,\ldots,d$. 
The first three functions are smooth (more precisely, they are $C^{\infty}(\Gamma)$),
whereas the Sobol G-function and $f_{C^0}$ have kink lines, hence are Lipschitz continuous but not $C^{1}(\Gamma)$,
and $f_{\mathrm{disc}}$ is discontinuous across the two segments $\{x_1=1/2, x_2 \leq 1/2\}$ and $\{x_1 \leq 1/2, x_2 = 1/2\}$.
As already mentioned, these last three functions are deliberately included as challenging cases for the globally supported polynomial sparse-grid approximations considered here. They provide a useful stress test for the bifidelity selection strategy to assess  whether the accuracy of the corresponding target sparse-grid approximation
can be recovered using only a fraction of the additional high-fidelity evaluations, even when the baseline and coarse sparse-grid approximations are not necessarily reliable throughout the entire domain.}

\revtext{We construct the bifidelity surrogate model considering both the absolute and relative indicators introduced in Equations~\eqref{eq:eta_abs} and~\eqref{eq:eta_indicator}; 
  for the relative indicator, we set $\varepsilon=10^{-5}$. We use the fractional budget approach to define $\eta_{\mathrm{cut}}$
  and prescribe the nominal fractions $q=0.1,0.2,\ldots,1$, see Equation \eqref{eq:Nq}.
  Since we also retain in $\mathcal{X}_{\mathrm{sel}}$ points with near-tied indicator values, as described in Section~\ref{sec:error-indicator},
  we report the computational cost of the resulting bifidelity surrogate models in terms of the percentage of candidate points actually evaluated, i.e.,
  \begin{equation}
    q_{\mathrm{eff}}=100 \times \frac{|\mathcal{X}_{\mathrm{sel}}|}{|\mathcal{X}_{\mathrm{cand}}|},
    \label{eq:effective-cost}
  \end{equation}
  Thus, $q_{\mathrm{eff}}=0\%$ corresponds to the baseline sparse-grid approximation, whereas $q_{\mathrm{eff}}=100\%$ corresponds to the complete target sparse-grid approximation.}

\revtext{The accuracy of the bifidelity surrogate models is evaluated as the RMSE over $M=2^{\lceil\log_2(1000d)\rceil}$ test points
  generated from a fixed scrambled Sobol sequence \cite{SOBOL196786}; 
  the same set of points is generated once for all tests to be performed for each value of $d$. 
  The reported RMSE is then further normalized with respect to the RMSE of the baseline sparse-grid approximation,
  such that the errors of the bifidelity surrogate models for $q=0$ are all equal to one.
}

\subsubsection{\revtext{Results across functions and dimensions}}

\begin{figure}[tp]
    \centering
    \includegraphics[width=\textwidth]{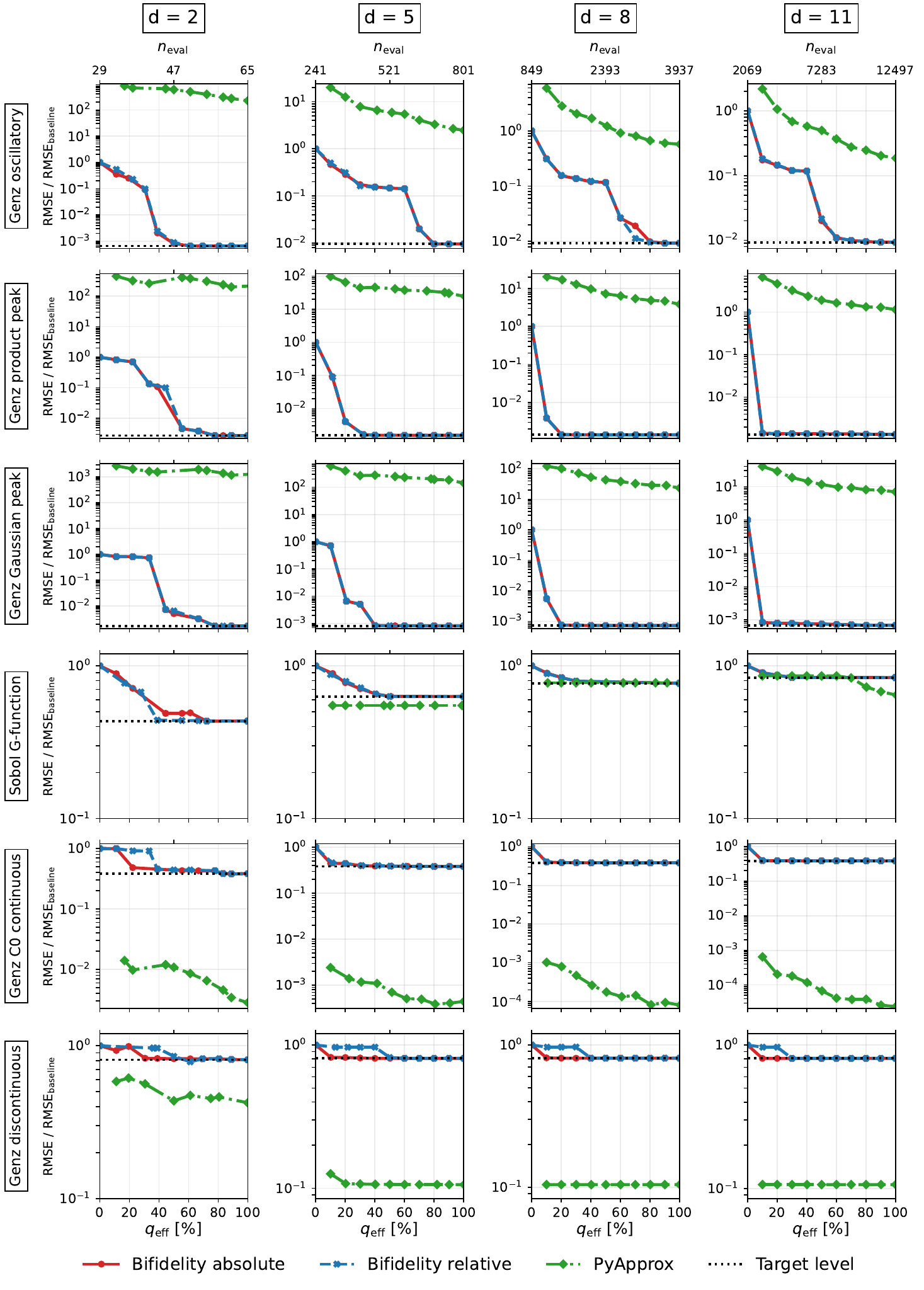}
    \caption{\revtext{Baseline-normalized RMSE versus $q_{\mathrm{eff}}$ for the test functions (rows) and dimensions (columns) discussed in Section \ref{sec:performance}, and the locally adaptive sparse-grid approximation discussed in Section \ref{sec:pyapprox-comparison}.
    For Sobol at $d=2$, the locally adaptive curve is not visible because its RMSE is below the plotted range (specifically, at machine precision, i.e, PyApprox interpolates exactly the function).}}
    \label{fig:cc-performance}
\end{figure}

 \revtext{Figure~\ref{fig:cc-performance} shows the results of the numerical tests previously described.
   The same figure also reports results obtained with locally adaptive sparse-grid approximations,
   that will be discussed separately in Section~\ref{sec:pyapprox-comparison}.}

 \revtext{As expected, the plots in Figure~\ref{fig:cc-performance} show that
   for smooth functions the target sparse-grid approximation (dotted horizontal line) achieves a substantial reduction of the RMSE,
   while for functions with limited regularity the RMSE does not decrease too much, even at the target level,
   due to the previously discussed difficulties encountered by sparse-grid approximations based on global polynomials when dealing with irregular functions.
   Regardless of this, the bifidelity construction approaches the accuracy of the prescribed target sparse-grid approximation
   using only a fraction of the candidate evaluations across all functions and dimensionalities.}

 \revtext{The largest savings are observed for the smooth peak functions: for the product-peak and Gaussian-peak cases at $d=8$ and $d=11$, target-level accuracy is reached at approximately $q_{\mathrm{eff}}=10$--$20\%$; this is justified by the fact that these peak functions are essentially constant on a large part of $\Gamma$, where the baseline sparse-grid approximation is already good enough. For the oscillatory function, the plateau is reached later, at $q_{\mathrm{eff}}=70$--$80\%$, corresponding to a reduction of about $20$--$30\%$ in the additional evaluations. These results confirm that a substantial fraction of the candidate points can be replaced by baseline predictions without significantly affecting the accuracy of the final surrogate model.} 


 \revtext{The results at $d=8$ and $d=11$ throughout all the test functions further show that the bifidelity surrogate model
   remains effective as the nominal dimension increases.
   For the smooth peak functions, the fraction of points required to approach the target even decreases with $d$,
which is expected since the portion of $\Gamma$ over which the peaks occur becomes smaller as $d$ increases.
In general, since $c_i\propto i^{-2}$ and $\sum_i c_i=1$, newly added dimensions have progressively smaller influence on the function, so the observed behavior shows that the proposed methodology can also exploit the dimensional anisotropy of the function to be approximated.}
 
\revtext{Finally, we observe that both the absolute and relative discrepancy indicators provide effective refinement. 
  The error curves of the corresponding bifidelity surrogate models are indeed nearly overlapping for the smooth peak functions, whereas some relatively minor differences emerge for the other test cases.} 

\subsubsection{\revtext{Comparison with locally adaptive sparse-grid approximations}}
\label{sec:pyapprox-comparison}

\revtext{We now compare the results of the bifidelity strategy with those of the locally adaptive piecewise-linear sparse-grid approximation implemented in PyApprox. To make results comparable, the RMSE of the locally adaptive sparse-grid approximation is also normalized with the RMSE of the baseline sparse-grid approximation, and in the software PyApprox we set the evaluation budget to be equal to that of the bifidelity surrogate models. Since PyApprox adds points in refinement batches, the final number of function evaluations might be slightly different from what was requested; we nonetheless plot the results as if they perfectly matched.}

\revtext{The comparison between the bifidelity and the locally adaptive strategy should be interpreted in light of their different approximation properties and goals. The bifidelity strategy aims at achieving the same accuracy as a target sparse-grid approximation based on global polynomials while avoiding evaluations that provide limited additional information, whereas locally adaptive sparse-grid approximations construct a piecewise-linear approximation that is locally refined in the vicinity of the local features of the function to be approximated.
  Thus, if such function is smooth ($f_{\mathrm{osc}}$, $f_{\mathrm{prod}}$, $f_{\mathrm{Gauss}}$),
  the target sparse-grid approximation with global polynomials attains significantly lower errors than the locally adaptive one, and the bifidelity strategy recovers such accuracy using only a subset of its candidate evaluations. 
  For the less regular functions ($f_{C^0}$, $f_{\mathrm{disc}}$, and $f_{\mathrm{Sobol}}$ for $d=2$),
locally adaptive \revtext{sparse-grid approximations} achieve lower errors than the target sparse-grid approximation (and its bifidelity surrogate model),
consistent with the advantage of locally supported basis functions near kinks and discontinuities.
In certain scenarios however (e.g., $f_{\mathrm{Sobol}}$ for $d=5,8,11$) the target sparse-grid approximation and the locally adaptive one achieve similar errors. }

\begin{figure}[t]
    \centering
    \includegraphics[width=0.5\textwidth]{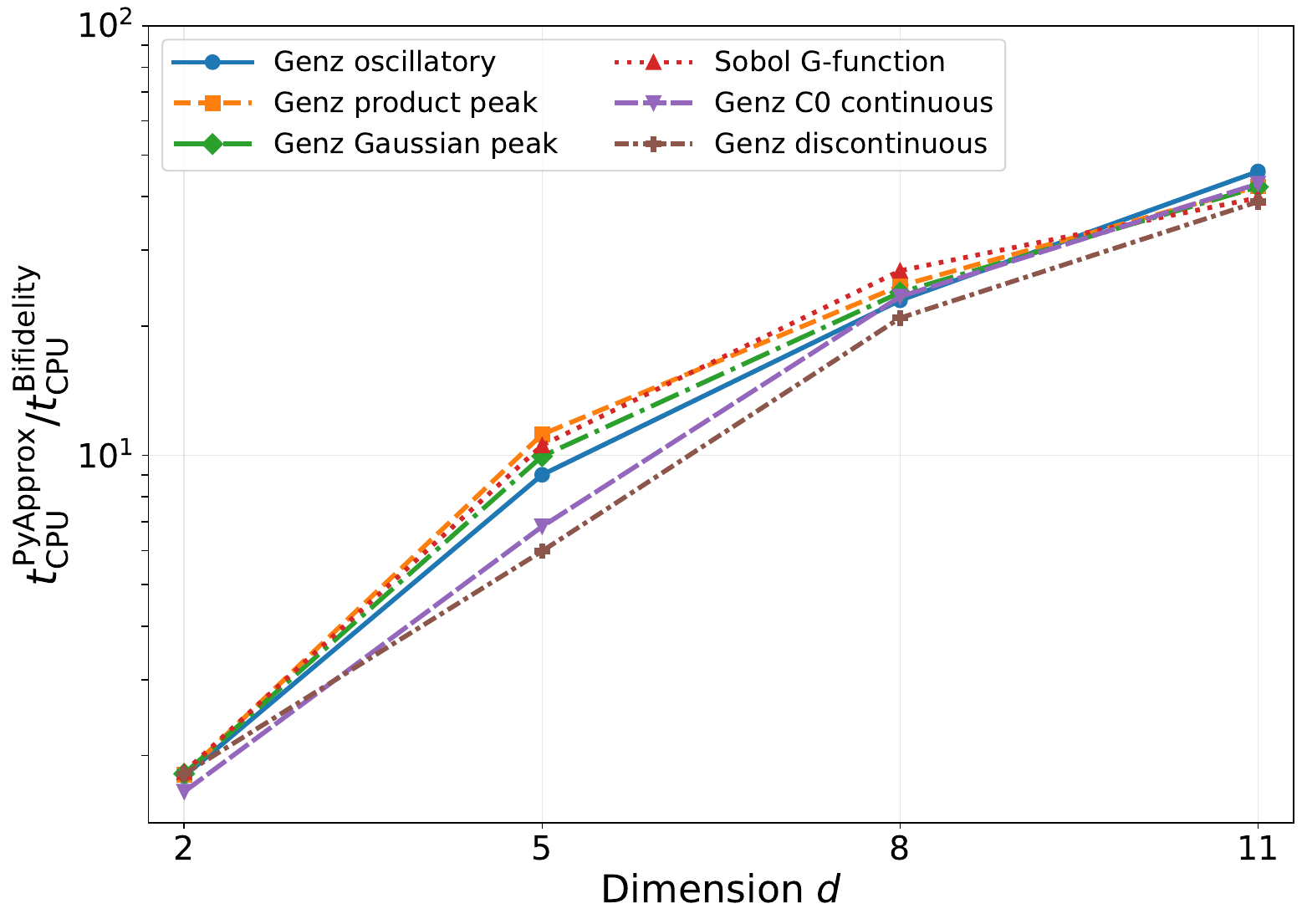}
    \caption{\revtext{Mean PyApprox-to-bifidelity CPU-time ratio as a function of the dimension $d$ for each test function in Section~\ref{sec:performance}. At each dimension, the ratio is averaged over the ten prescribed $q$ values.}}
    \label{fig:cpu-performance}
\end{figure}

\revtext{The two approaches should also be compared in terms of other metrics, such as 
  the algorithmic overhead associated with managing their respective data structures changes as $d$ grows. To this end, we report in Figure \ref{fig:cpu-performance} 
  the mean over the ten prescribed values of $q$ (see Section \ref{sec:performance-setup}) 
  of the ratio of the computational times of the locally adaptive and bifidelity approaches, $t_{\mathrm{CPU}}^{\mathrm{PyApprox}}/t_{\mathrm{CPU}}^{\mathrm{bifidelity}}$,
  as $d$ increases. The bifidelity strategy used here employs the absolute discrepancy indicator.  
  The timings include surrogate model construction or refinement and prediction at the test points, 
  i.e., both timings entail essentially the same number of function evaluations. 
  Because the benchmark functions are inexpensive to evaluate, 
  these CPU times mainly quantify the algorithmic overhead due to managing the data structures. 
  Their ratio increases from approximately two at $d=2$ to several tens at $d=11$, highlighting
  the significant relative degradation of performance of the locally adaptive algorithm, which is well-known in literature. 
  Of course, this difference in CPU times becomes irrelevant
in applications dominated by expensive high-fidelity evaluations,
where the main computational advantage of the bifidelity approach derives from its ability to exploit whatever smoothness is available to achieve the same accuracy as an approximation based on global polynomials while requiring fewer model evaluations.}

\subsection{Application to \revtext{a hydrogen-fueled perforated-burner} case study}
\label{sec:experiment}

The primary objective of the proposed bifidelity methodology is to enable efficient and accurate surrogate modeling of complex engineering systems, for which direct evaluations of the full numerical model are computationally expensive. To demonstrate its practical applicability, we consider a representative case study concerning the design of a hydrogen-fueled perforated burner. 
In such systems, flashback phenomena pose critical risks due to hydrogen’s high flame speed and diffusivity. It is therefore essential to systematically assess how the flashback behavior is influenced by the geometry of the burner,
to optimize the design of such devices while guaranteeing safety and control strategies, as well as to select configurations of experimental interest (see also \cite{FRUZZA2026mapping} for a related sensitivity analysis study). With these objectives in mind, our goal here is to derive a surrogate model of the flashback velocity as a function of the geometrical design parameters of the burner.

\begin{figure*}[t]
    \centering
    \includegraphics[width=0.8\linewidth]{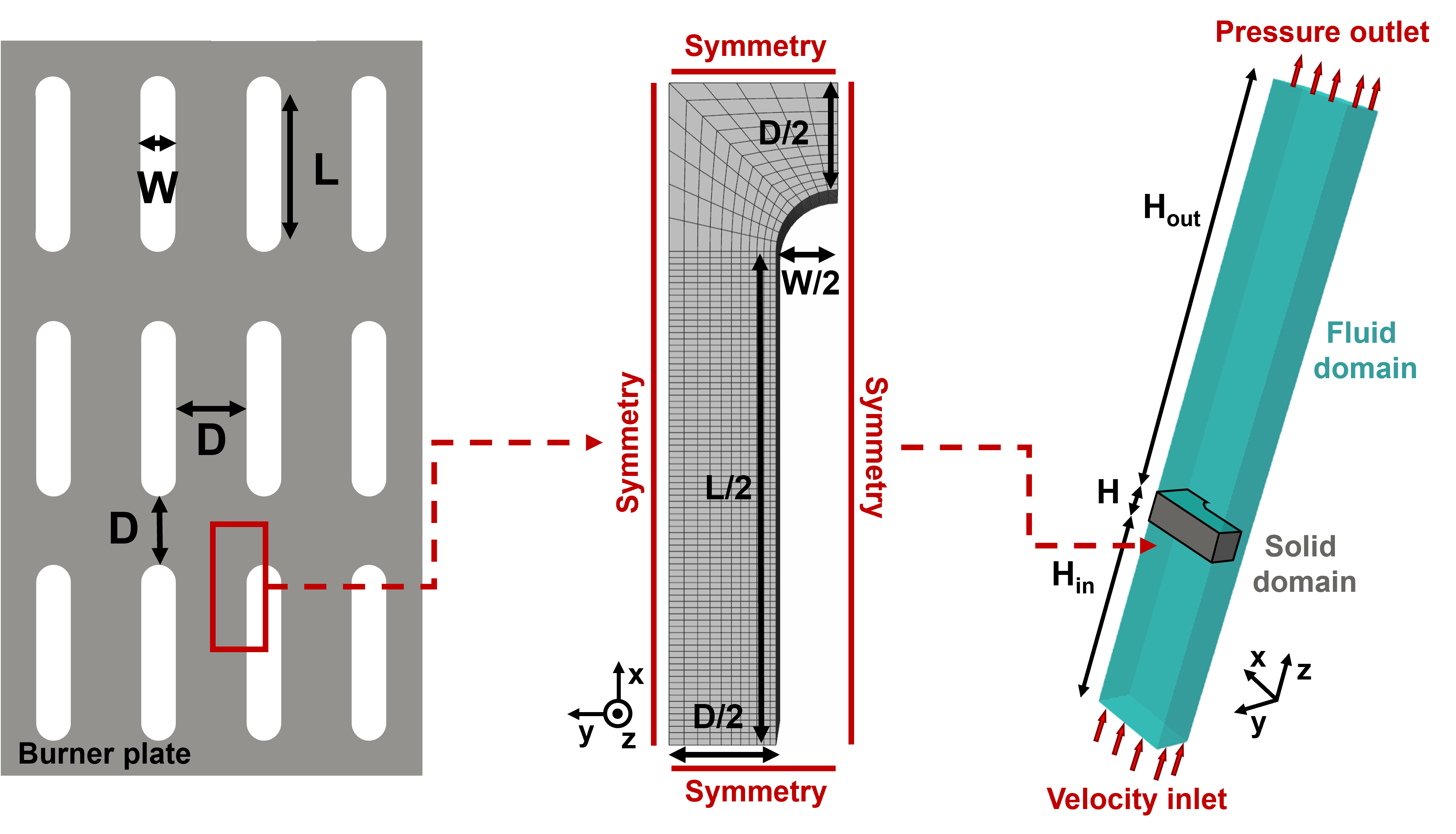}
    \caption{Left panel: Slit pattern on the burner plate, highlighting the section of the computational domain. Center panel: Slit geometry (boundary conditions in red) and solid plate mesh. Right panel: Computational domain (fluid zone in light blue, solid zone in gray, boundary conditions in red). Reproduced from~\cite{FRUZZA2025_CF}.}
    \label{fig:domain}
\end{figure*}

In detail, we consider a three-dimensional segment of a perforated stainless steel burner plate with a single slit, parameterized by four geometric variables: slit length ($L$), slit width ($W$), inter-slit spacing ($D$), and plate thickness ($H$), as shown in Figure~\ref{fig:domain}. 
The parameters vary within the following ranges:
$L \in [0,\,4]~\text{mm}$,
$W \in [0.4,\,0.8]~\text{mm}$,
$D \in [0.3,\,1.7]~\text{mm}$,
and $H \in [0.3,\,0.9]~\text{mm}$.
In this computational domain, we consider the flow of a mixture of air and hydrogen, 
modeled by steady, laminar, compressible Navier--Stokes equations, 
coupled with diffusion-transport-reaction equations for the chemical species, 
together with an ideal-gas equation of state. 
The simulations consider a hydrogen–air mixture at equivalence ratio $\phi = 0.6$, with uniform inlet conditions 
($T_u = 300~\mathrm{K}$) 
and an ambient pressure outlet ($p = 1~\mathrm{atm}$). 
For all cases investigated, the Reynolds number of the resulting jet remains within the laminar regime and no turbulence modeling is employed. Thermal coupling between the fluid and solid domains is enforced through a conjugate heat transfer formulation, 
while chemical kinetics are described using a reduced mechanism comprising 9 species and 22 reactions~\cite{kee}. 

The governing equations are discretized on a structured three-dimensional mesh with local refinement in the reaction region. The characteristic cell size in the flame zone is approximately $\Delta x \simeq 25~\mu\mathrm{m}$, corresponding to about $\delta_F/13$, where $\delta_F$ denotes the unstretched one-dimensional thermal flame thickness. Depending on the geometric configuration, the total number of control volumes ranges from approximately $3\times10^{5}$ to $3\times10^{6}$. This resolution was selected based on grid-independence analyses reported in previous studies~\cite{FRUZZA2023,fruzzaUQ,FRUZZA_PROCI,FRUZZA2025_CF}. All simulations are performed using \textit{ANSYS Fluent 24.2}. 

The critical flashback condition is determined through a sequence of steady reacting-flow simulations. For a given burner geometry, a stable flame solution is first established at a sufficiently high inlet velocity, ensuring that flashback does not occur. The inlet velocity is then progressively reduced, and successive steady-state simulations are performed until the solver can no longer converge to a stable flame solution. This loss of convergence is interpreted as the onset of flashback. The corresponding critical flashback velocity is defined as the cold-flow bulk velocity at the slit entry,
\[
  V_\mathrm{FB} = \frac{A_\mathrm{tot}}{A_\mathrm{slit}}\, V_\mathrm{in} = \frac{V_\mathrm{in}}{\psi},
\]
where $V_\mathrm{in}$ is the imposed inlet velocity and $\psi = A_\mathrm{slit}/A_\mathrm{tot}$ denotes the burner porosity. This definition provides an intrinsic measure of the configuration resistance to flashback, independent of trivial geometric scaling. Further details on the numerical procedure and its validation are reported in~\cite{FRUZZA2023,fruzzaUQ,FRUZZA_PROCI,FRUZZA2025_CF}.
Each evaluation of $V_\mathrm{FB}$ thus requires multiple fully coupled three-dimensional reacting-flow simulations with detailed chemistry and conjugate heat transfer. As a result, a single high-fidelity evaluation typically entails wall-clock times of several days up to approximately one week on $\mathcal{O}(10^2)$ parallel processors. This computational burden renders exhaustive sampling of the parametric domain impractical and provides the primary motivation for the (bifidelity) surrogate modeling strategy introduced in this work. 

In a previous study~\cite{FRUZZA2026mapping}, an earlier version of the bifidelity strategy described in the current work was proposed
to refine the baseline sparse-grid approximation (obtained also in this case for $w=2$) in the region $L \approx 0$ of the parametric domain;
specifically, the chosen set $\mathcal{X}_{\mathrm{sel}}$ consisted of 33 points.
Crucially, however, this decision was guided by expert judgment, based on prior experience and knowledge/physical intuition about the problem at hand;
in the following, we refer to this construction as ``human-based bifidelity approach''. 
In the current study, we aim to make the bifidelity strategy entirely algorithmic
(in the following, ``algorithmic bifidelity approach''), by using the methodology presented above. 

\begin{figure}[t]
    \centering
    \includegraphics[width=0.40\textwidth]{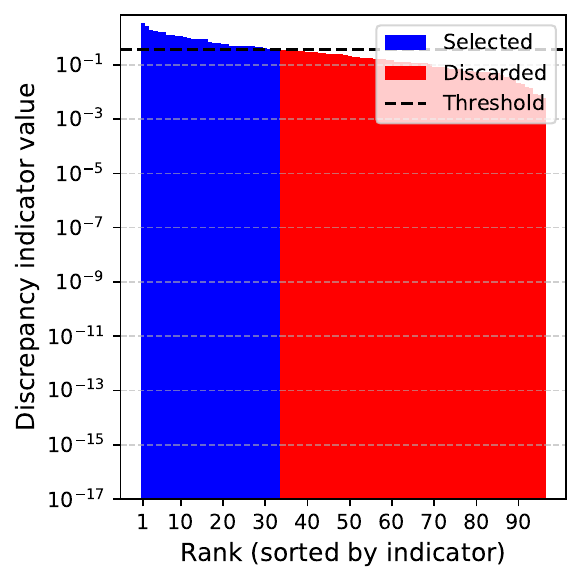}
    \includegraphics[width=0.48\textwidth]{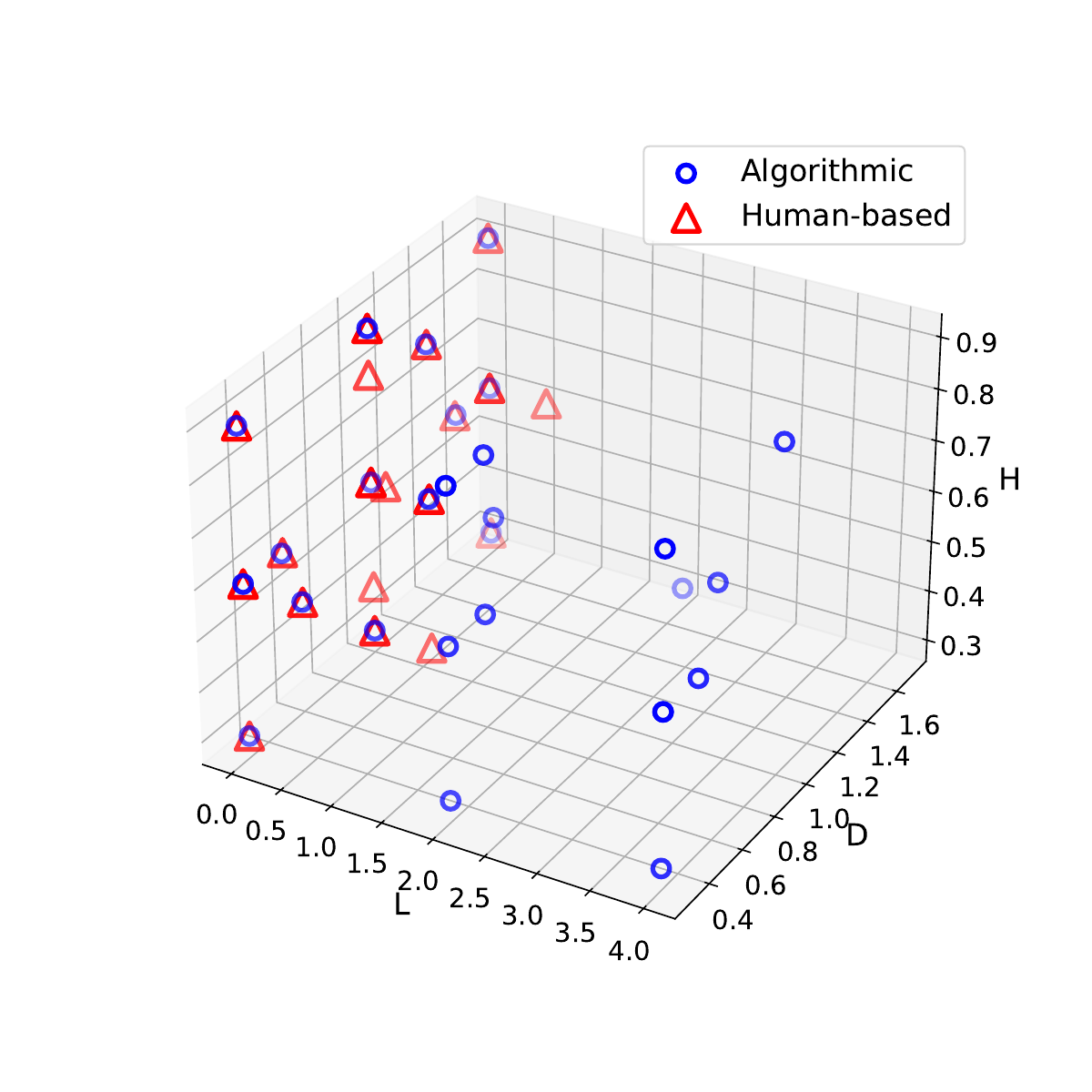}
    \caption{Discrepancy indicator for collocation points in $\mathcal{X}_{\mathrm{cand}}$  for the burner problem,
      sorted decreasingly (left) and $\mathcal{X}_{\mathrm{sel}}$ for the human-based and algorithmic strategies (right)\revtext{.}}
    \label{fig:flames-discrepancy}
\end{figure}
The two scalar output quantities of interest are flashback velocity and maximum burner temperature. Thus, the refinement process can be guided by either quantity. We observed that the resulting performance exhibits no significant differences depending on the variable chosen to drive the refinement, so in what follows, we refer to the process using the flashback velocity as the guiding quantity. Figure~\ref{fig:flames-discrepancy} (left) presents the discrepancy indicator computed for the points in $\mathcal{X}_{\mathrm{cand}}$, ranked in decreasing order of importance.
To ensure a fair comparison between the human-based approach proposed in \cite{FRUZZA2026mapping} and the automatic bifidelity strategy, the computational budget in this study was fixed at 33 points.
The sets $\mathcal{X}_{\mathrm{sel}}$ selected by the human-based and the algorithmic approaches are then shown in 
Figure~\ref{fig:flames-discrepancy} (right), in the three-dimensional $\left(L, D, H\right)$ space at $W=0.6~\text{mm}$.
While the two sets are not identical, the bifidelity surrogate model successfully recovers the majority of the points identified by the human-based approach, particularly in regions where the function exhibits strong gradients or localized features. This overlap confirms the effectiveness of the discrepancy indicator in targeting the most informative regions of the domain. In addition to this basic agreement, the bifidelity strategy introduces a few supplementary points in more peripheral or less active areas of the parametric domain. 
Overall, the comparison highlights the capacity of the algorithmic strategy to emulate expert-driven refinement while offering a more systematic and scalable alternative.

\begin{figure}[t]
    \centering
    \includegraphics[width=0.48\textwidth]{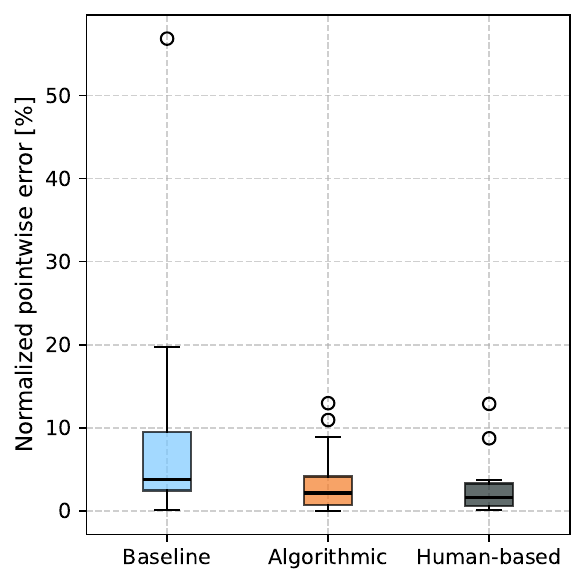}
    \caption{Boxplots of the percentage normalized error for the baseline \revtext{sparse-grid approximation},
      and human-based and algorithmic bifidelity surrogate \revtext{models} for the flashback velocity.}
    \label{fig:flames-histogram}
\end{figure}

Figure \ref{fig:flames-histogram} shows 
the boxplots of the percentage normalized pointwise error for the baseline \revtext{sparse-grid approximation},
the human-based bifidelity \revtext{surrogate model}, and the algorithmic bifidelity \revtext{surrogate model}. 
Since no reference values are available in closed-form for this {computational case study},
the comparison is performed against 17 additional simulations, conducted at parameter values selected to be as far as possible from the
{collocation} points used to build the surrogate models. 
The results clearly demonstrate that the algorithmic approach achieves a substantial improvement over the baseline \revtext{sparse-grid approximation}
and confirms its effectiveness in capturing the underlying physical behavior with minimal computational cost.
Importantly, it also achieves error distributions closely aligned with those of the human-based approach.
The small difference should be weighed against the fact that such expert-driven refinement is feasible only when deep physical insight and domain-specific experience are available. By contrast, the algorithmic strategy requires no prior knowledge and is fully automated, making its near-expert performance particularly valuable and naturally scalable.

\begin{figure}[t]
    \centering
    \includegraphics[width=0.9\linewidth]{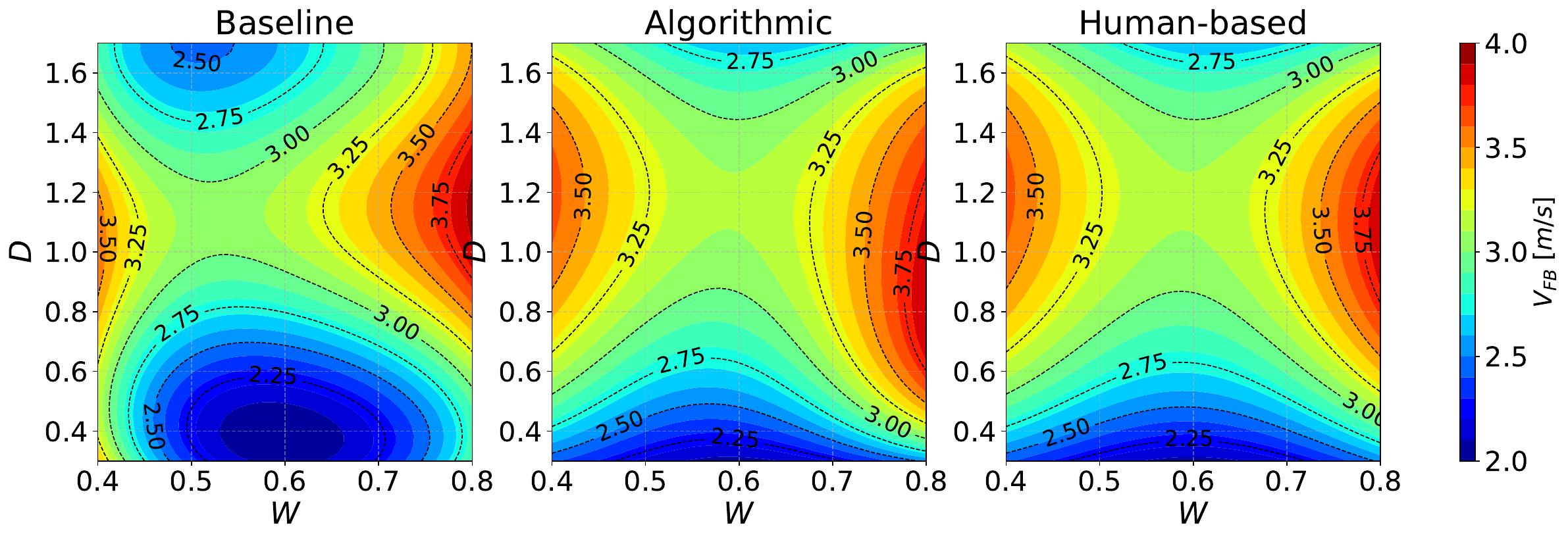}
    \caption{Comparison of the  baseline \revtext{sparse-grid approximation},
      the human-based bifidelity surrogate model, and the {algorithmic} bifidelity surrogate model
      for the flashback velocity in our case study at $L=0.0$ and $H=0.6$.}
    \label{fig:flames-response1}
\end{figure}

Finally, Figure \ref{fig:flames-response1} compares the response surfaces for the flashback velocity predicted by the baseline \revtext{sparse-grid approximation},
the human-based bifidelity surrogate model, and the {algorithmic} bifidelity surrogate model.
As expected from the error boxplots, the human-based and algorithmic bifidelity surrogate models produce nearly identical predictions,
exhibiting a sharper representation of the velocity gradient across the domain,
especially in regions where the baseline \revtext{sparse-grid approximation} tends to oversmooth the underlying physical behavior.

\section{Conclusions}

In this work, we introduce a novel bifidelity framework for constructing bifidelity \revtext{surrogate models},
which can lead to significant computational savings, \revtext{especially in complex applications of practical interest
  where each evaluation of the high-fidelity model entails a high computational cost.}
The core of our contribution is a zero-cost discrepancy indicator that \revtext{leverages} two nested,
pre-existing \revtext{sparse-grid approximations} to obtain an efficient a-priori ranking of candidate points
\revtext{in the parametric domain} where the expensive \revtext{model} can be evaluated \revtext{to refine the sparse-grid approximation}.
\revtext{This enables the construction of a more accurate bifidelity surrogate model by allocating the computational budget exclusively to the most impactful regions of the parametric domain.}

The effectiveness of the proposed methodology was validated through a series of numerical investigations. 
First, its robustness and efficacy \revtext{were} demonstrated on a set of benchmark functions. 
The results consistently showed that the bifidelity surrogate model achieves an accuracy closely approaching that of a target sparse-grid approximation, while requiring only a fraction of the computational budget. This confirms that the proposed zero-cost indicator is highly effective at identifying the most salient points for function evaluation. 

The practical applicability of the framework was then demonstrated on a computational dataset from a hydrogen-fueled perforated burner problem. In this real-world scenario, our method yielded excellent results, comparable to those obtained from a specialized, human-based refinement of the surrogate model. A remarkable finding was that our algorithm automatically identified the majority of the refinement points that had been manually selected in the human-based approach based on specific knowledge. 

Future work will focus on applying this framework to a broader range of complex engineering problems, particularly in the context of optimization and uncertainty quantification. Further research could also explore the formulation of more sophisticated indicators and the integration of this adaptive strategy within multi-fidelity modeling workflows.

\paragraph{Acknowledgments}
This work has been supported by the ICSC-Centro Nazionale di Ricerca in High Performance Computing, Big Data, and Quantum Computing, funded by European Union - NextGenerationEU.
Lorenzo Tamellini has been partially supported by the project 202222PACR ``Numerical approximation of uncertainty quantiﬁcation problems for PDEs by multi-ﬁdelity methods (UQ-FLY)''
and is a member of the Gruppo Nazionale Calcolo Scientifico-Istituto Nazionale di Alta Matematica (GNCS-INdAM).

\paragraph{Data Availability Statement}
The Python toolbox developed for the algorithmic bifidelity approach is open‑source and available at \url{https://github.com/cfdlab-unipi/surrogate-informed-sparse-grid-tool}. It is released under the MIT license, allowing unrestricted use, modification, and redistribution. The toolbox implements in Python some of the basic functionalities of the Sparse Grids Matlab Kit \cite{piazzola2024algorithm}
and extends it by incorporating the algorithmic bifidelity logic introduced in this work, providing a fully Python‑based framework for surrogate model construction and analysis. It supports scalar outputs, time-series data, and vector-valued quantities. All scripts required to reproduce the figures and results presented in this work are included in the repository.

\paragraph{Credit Statement} 
\textbf{Matteo Rosellini:} Conceptualization, implementation and data curation, Writing -  first draft; 
\textbf{Filippo Fruzza:} Conceptualization, implementation and data curation, Writing - first draft; 
\textbf{Alessandro Mariotti:} Methodology, Writing – review \& editing. 
\textbf{Maria Vittoria Salvetti:} Methodology, Conceptualization, Writing – review \& editing. 
\textbf{Lorenzo Tamellini:} Methodology, Conceptualization, Writing – review \& editing.


\bibliographystyle{unsrt}
\bibliography{references}

\end{document}